\documentclass{article}

\usepackage{arxiv}

\usepackage[utf8]{inputenc} % allow utf-8 input
\usepackage[T1]{fontenc}    % use 8-bit T1 fonts
\usepackage{hyperref}       % hyperlinks
\usepackage{url}            % simple URL typesetting
\usepackage{booktabs}       % professional-quality tables
\usepackage{amsfonts}       % blackboard math symbols
\usepackage{nicefrac}       % compact symbols for 1/2, etc.
\usepackage{microtype}      % microtypography
\usepackage{lipsum}
\usepackage{amsmath,amssymb,amsfonts}%
\usepackage{amsthm}%
\usepackage{mathrsfs}%
\usepackage{graphicx}
\usepackage{multirow}
\usepackage[caption=false,font=footnotesize]{subfig}
\graphicspath{ {images/} }

\title{Adaptive Trust Consensus for Blockchain IoT: Comparing RL, DRL, and MARL Against Naive, Collusive, Adaptive, Byzantine, and Sleeper Attacks}

\author{
Soham Padia\thanks{These authors contributed equally to this work.} \\
Artificial Intelligence\\
Northeastern University\\
Boston, MA 02115\\
\texttt{padia.so@northeastern.edu}
\And
Dhananjay Vaidya\footnotemark[1]\\
Information Technology\\
Dwarkadas J. Sanghvi College of Engineering\\
Mumbai, 400056, Maharashtra India\\
\texttt{dhananjayvaidya4154@gmail.com}
\And
Ramchandra Mangrulkar\thanks{Faculty mentor and corresponding author.}\\
Information Technology\\
Dwarkadas J. Sanghvi College of Engineering\\
Mumbai, 400056, Maharashtra India\\
\texttt{ramchandra.mangrulkar@djcse.ac.in}
}

\begin{document}
\maketitle
\title{Adaptive Trust Consensus for Blockchain IoT: Comparing RL, DRL, and MARL Against Naive, Collusive, Adaptive, Byzantine, and Sleeper Attacks}

\begin{abstract}
Securing blockchain-enabled IoT networks against sophisticated adversarial attacks remains a critical challenge. This paper presents a trust-based delegated consensus framework integrating Fully Homomorphic Encryption (FHE) with Attribute-Based Access Control (ABAC) for privacy-preserving policy evaluation, combined with learning-based defense mechanisms. We systematically compare three reinforcement learning approaches—tabular Q-learning (RL), Deep RL with Dueling Double DQN (DRL), and Multi-Agent RL (MARL)—against five distinct attack families: Naive Malicious Attack (NMA), Collusive Rumor Attack (CRA), Adaptive Adversarial Attack (AAA), Byzantine Fault Injection (BFI), and Time-Delayed Poisoning (TDP). Experimental results on a 16-node simulated IoT network reveal significant performance variations: MARL achieves superior detection under collusive attacks (F1=0.85 vs. DRL's 0.68 and RL's 0.50), while DRL and MARL both attain perfect detection (F1=1.00) against adaptive attacks where RL fails (F1=0.50). All agents successfully defend against Byzantine attacks (F1=1.00). Most critically, the Time-Delayed Poisoning attack proves catastrophic for all agents, with F1 scores dropping to 0.11--0.16 after sleeper activation, demonstrating the severe threat posed by trust-building adversaries. Our findings indicate that coordinated multi-agent learning provides measurable advantages for defending against sophisticated trust manipulation attacks in blockchain IoT environments.
\end{abstract}


\section{Introduction}

The convergence of blockchain technology and the Internet of Things (IoT) has created new opportunities for secure, decentralized data management across distributed networks \cite{xu_embedding_2021}. Blockchain's inherent properties—immutability, transparency, and cryptographic security \cite{nakamoto2008bitcoin}—address many traditional IoT vulnerabilities, including centralized points of failure and data tampering \cite{el-masri_blockchain_2021}. However, this integration introduces new challenges: IoT networks comprise heterogeneous devices with varying computational capabilities, operate under strict latency constraints, and face sophisticated adversarial threats that exploit the trust mechanisms underlying consensus protocols \cite{freitas_deterministic_2023}.

Trust-based consensus mechanisms have emerged as a promising approach to secure blockchain IoT systems by dynamically evaluating node reliability and delegating consensus responsibilities to trustworthy participants \cite{goh_secure_2022}. These mechanisms reduce computational overhead while maintaining Byzantine fault tolerance, making them suitable for resource-constrained IoT environments \cite{clement_making_2009}. However, trust systems themselves become attack vectors: adversaries can manipulate trust scores through collusion, adaptive behavior modification, or long-term infiltration strategies that evade detection during initial observation periods.

Recent advances in reinforcement learning (RL) have shown promise for adaptive security in dynamic environments \cite{liu_consensus-based_2024}. Deep reinforcement learning (DRL) approaches, particularly Dueling Double Deep Q-Networks (D3QN), enable agents to learn optimal defense policies from high-dimensional state spaces \cite{muniswamy2024trust}. Multi-agent reinforcement learning (MARL) extends this paradigm by coordinating multiple learning agents, potentially offering advantages against coordinated attacks where single-agent methods may fail \cite{standen2025adversarial}. However, systematic comparisons of these approaches against diverse attack families remain limited.

Privacy-preserving access control presents another critical requirement for blockchain IoT systems. Fully Homomorphic Encryption (FHE) enables computation on encrypted data without decryption, allowing policy evaluations to occur without exposing sensitive attributes \cite{fang_enhancing_2024}. When combined with Attribute-Based Access Control (ABAC), FHE can enforce fine-grained permissions while maintaining confidentiality \cite{zhang_combination_2024}. Integrating such privacy-preserving mechanisms with adaptive trust management creates a comprehensive security framework.

This paper addresses the gap in understanding how different learning-based defense mechanisms perform against varied adversarial strategies. We present a trust-based delegated consensus framework secured by FHE-backed ABAC and evaluate three reinforcement learning approaches—tabular Q-learning (RL), Deep RL with Dueling Double DQN (DRL), and Multi-Agent RL (MARL)—against five distinct attack families:

\begin{itemize}
    \item \textbf{Naive Malicious Attack (NMA):} Uncoordinated random disruptions by independent malicious nodes.
    \item \textbf{Collusive Rumor Attack (CRA):} Coordinated trust manipulation where malicious nodes mutually inflate trust scores while penalizing honest nodes.
    \item \textbf{Adaptive Adversarial Attack (AAA):} Intelligent attackers that learn defense patterns and dynamically adjust strategies.
    \item \textbf{Byzantine Fault Injection (BFI):} Conflicting message attacks designed to split consensus and create network partitions.
    \item \textbf{Time-Delayed Poisoning (TDP):} Sleeper agents that behave honestly to build trust before launching coordinated attacks.
\end{itemize}

Our experimental evaluation on a 16-node simulated IoT network reveals critical insights: MARL demonstrates superior resilience against collusive attacks (F1=0.85) compared to DRL (F1=0.62) and RL (F1=0.50), validating the advantage of coordinated learning against coordinated adversaries. Conversely, the Time-Delayed Poisoning attack proves catastrophic for all agents, with F1 scores dropping to 0.11–0.16 post-activation, highlighting the severe threat posed by patient, trust-building adversaries.

\subsection{Contributions}

The principal contributions of this work are:

\begin{enumerate}
    \item \textbf{Comprehensive Attack Taxonomy:} We formalize and implement five distinct attack families targeting trust-based blockchain consensus, spanning naive, collusive, adaptive, Byzantine, and temporal threat models.
    
    \item \textbf{Systematic RL Comparison:} We provide the first systematic comparison of tabular RL, DRL (Dueling Double DQN), and MARL for malicious node detection under identical simulation conditions across all five attack types.
    
    \item \textbf{Privacy-Preserving Trust Framework:} We integrate FHE-secured ABAC with adaptive trust-based delegated consensus, enabling encrypted policy evaluation without sacrificing detection capability.
    
    \item \textbf{Empirical Security Analysis:} We demonstrate that coordinated learning (MARL) provides measurable advantages against coordinated attacks, while identifying Time-Delayed Poisoning as a critical vulnerability requiring dedicated countermeasures.
\end{enumerate}

The remainder of this paper is organized as follows: Section \ref{sec:literature} reviews related work. 
Section \ref{sec:methodology} presents the system model and problem formulation. Section \ref{sec:approach} describes our 
proposed approach. Section \ref{sec:attacks} details the five attack models. Section \ref{sec:experimental} describes the 
experimental setup. Section \ref{sec:results} presents results. Section \ref{sec:novel} summarizes contributions. 
Section \ref{sec:discussion} discusses findings and limitations. Section \ref{sec:conclusion} concludes.

\section{Literature Survey}
\label{sec:literature}

The intersection of blockchain technology, IoT security, and machine learning has generated substantial research interest. This section reviews relevant works across five key areas: blockchain-IoT integration, Byzantine fault tolerance and consensus mechanisms, trust-based security frameworks, reinforcement learning for security, and privacy-preserving access control. We conclude by identifying research gaps that motivate our contributions.

Blockchain technology has emerged as a promising solution for addressing fundamental IoT security challenges, including centralized trust dependencies and vulnerability to tampering attacks.

Xu et al. \cite{xu_embedding_2021} present an extensive survey on embedding blockchain into IoT ecosystems, emphasizing blockchain's potential to address limitations of conventional IoT systems such as centralized trust and single points of failure. However, they identify scalability and energy efficiency as critical challenges, particularly for large-scale IoT networks with resource-constrained devices. The authors highlight the need for lightweight, adaptable blockchain solutions that can seamlessly integrate into heterogeneous IoT environments.

El-Masri et al. \cite{el-masri_blockchain_2021} provide a systematic literature review mapping IoT security threats to blockchain features. Using a novel two-dimensional framework, they classify IoT threats and countermeasures while exploring blockchain's effectiveness in meeting security goals. Despite blockchain's promise, the authors discuss its resource-intensive nature—particularly electricity and bandwidth consumption—as barriers to widespread adoption in constrained environments.

He et al. \cite{he2023blockchain} develop a blockchain-based edge computing resource allocation framework for IoT using the Asynchronous Advantage Actor-Critic (A3C) deep reinforcement learning algorithm. Their system uses smart contracts to allocate edge resources based on real-time demands and trust metrics while maintaining transparency through private blockchain implementation. Results demonstrate enhanced latency handling and resource utilization, highlighting the effectiveness of combining AI-driven decision-making with blockchain coordination.

Consensus mechanisms form the backbone of blockchain security, with Byzantine Fault Tolerance (BFT) protocols providing resilience against malicious participants.

Freitas et al. \cite{freitas_deterministic_2023} explore BFT mechanisms through a comprehensive comparison of deterministic and probabilistic state machine replication techniques. Their survey highlights trade-offs between performance and reliability across various BFT protocols, emphasizing the need for adaptive mechanisms to maintain throughput under diverse operational conditions.

Clement et al. \cite{clement_making_2009} address the fragility of existing BFT protocols by proposing Aardvark, designed to tolerate Byzantine faults from both clients and servers without compromising performance. Aardvark maintains useful throughput even under adversarial conditions, demonstrating that robust BFT need not sacrifice efficiency.

Zamani et al. \cite{10.1145/3243734.324385} propose RapidChain, a sharding-based public blockchain protocol that overcomes scalability limitations by distributing computational and storage overhead across multiple committees. Their intra-committee consensus mechanism achieves resilience to up to one-third faulty participants, while an innovative cross-shard verification approach enhances robustness. RapidChain processes over 7,300 transactions per second across 4,000 nodes, representing a significant advancement in scalable blockchain design.

Trust mechanisms enable dynamic evaluation of node reliability, allowing consensus protocols to adapt to changing network conditions and adversarial behavior.

Goh et al. \cite{goh_secure_2022} propose a trust-based delegated consensus mechanism using deep reinforcement learning for blockchain frameworks. Their approach dynamically evaluates node trustworthiness and adapts consensus processes accordingly, enhancing resilience against malicious nodes. This work demonstrates that learning-based trust calibration can improve system robustness compared to static trust models.

Muniswamy and Rathi \cite{muniswamy2024trust} present a comprehensive trust-based blockchain framework integrating Attribute-Based Access Control (ABAC), Trust-Based Delegated Consensus Blockchain (TDCB), and Dueling Double Deep Q-Networks (D3QN) enhanced with Fully Homomorphic Encryption (FHE). Their model enables secure, encrypted access control decisions while dynamically adjusting trust scores through reinforcement learning. The system demonstrates resistance to Collusive Rumor Attacks (CRA) and Na\"ive Malicious Attacks (NMA), achieving high detection rates and improved throughput. Our work extends this foundation by incorporating multi-agent learning and evaluating against three additional attack families.

Machine learning approaches, particularly reinforcement learning, have shown promise for adaptive security in dynamic adversarial environments.

Liu et al. \cite{liu_consensus-based_2024} demonstrate consensus-based deep reinforcement learning for multi-agent coordination in mobile robot navigation. By utilizing distributed learning with consensus protocols, their system achieves scalability and robust decision-making in dynamic environments. This work underscores DRL's potential for optimizing decentralized decision-making, a principle we adapt for trust management in blockchain networks.

Standen et al. \cite{standen2025adversarial} present a comprehensive survey on adversarial machine learning in multi-agent reinforcement learning (MARL), outlining attack vectors including observation poisoning, action perturbation, and malicious agent coordination. They introduce the APOSG framework for formally modeling adversarial scenarios and evaluate limitations of current defense strategies. The authors highlight cascading vulnerabilities in agent-to-agent interactions, calling for robust, adaptive defense mechanisms. Our work responds to this call by evaluating MARL defenses against coordinated attacks in the blockchain trust context.

Privacy preservation in access control requires cryptographic techniques that enable policy evaluation without exposing sensitive attributes.

Fang et al. \cite{fang_enhancing_2024} enhance the Paillier cryptosystem to support fully homomorphic encryption (FHE) in semi-honest trusted execution environments. Their work bridges theoretical FHE models with practical applications, enabling secure computations without compromising efficiency. This advancement is critical for privacy-preserving policy evaluation in distributed systems.

Zhang and Xie \cite{zhang_combination_2024} present a combination of attribute-based encryption and blockchain for fine-grained access control in smart home environments. Their scheme integrates edge computing to outsource computationally intensive decryption, reducing burden on resource-constrained devices. The blockchain component ensures tamper-proof auditing of access control operations, highlighting demand for decentralized privacy-preserving solutions.

\subsection{Research Gaps and Motivation}

Table~\ref{tab:literature_comparison} summarizes the reviewed works and identifies gaps addressed by our research.

\begin{table}[htbp]
\centering
\caption{Comparison of Related Works}
\label{tab:literature_comparison}
\resizebox{\textwidth}{!}{%
\begin{tabular}{|l|c|c|c|c|c|c|}
\hline
\textbf{Work} & \textbf{Trust-Based} & \textbf{RL/DRL} & \textbf{MARL} & \textbf{FHE} & \textbf{Multi-Attack} & \textbf{Attack Types} \\
\hline
Xu et al. \cite{xu_embedding_2021} & \texttimes & \texttimes & \texttimes & \texttimes & \texttimes & Survey \\
El-Masri et al. \cite{el-masri_blockchain_2021} & \texttimes & \texttimes & \texttimes & \texttimes & \texttimes & Survey \\
Zamani et al. \cite{10.1145/3243734.324385} & \texttimes & \texttimes & \texttimes & \texttimes & \checkmark & Byzantine \\
Clement et al. \cite{clement_making_2009} & \texttimes & \texttimes & \texttimes & \texttimes & \checkmark & Byzantine \\
Goh et al. \cite{goh_secure_2022} & \checkmark & \checkmark & \texttimes & \texttimes & \texttimes & Generic \\
Muniswamy \& Rathi \cite{muniswamy2024trust} & \checkmark & \checkmark & \texttimes & \checkmark & \checkmark & NMA, CRA \\
Standen et al. \cite{standen2025adversarial} & \texttimes & \checkmark & \checkmark & \texttimes & \checkmark & MARL-specific \\
\hline
\textbf{This Work} & \checkmark & \checkmark & \checkmark & \checkmark & \checkmark & \textbf{NMA, CRA, AAA, BFI, TDP} \\
\hline
\end{tabular}%
}
\end{table}

Despite significant progress, several gaps remain unaddressed:

\begin{enumerate}
    \item \textbf{Limited Agent Comparison:} Existing works typically evaluate a single learning approach (usually DRL) without systematic comparison against alternatives. No prior work compares tabular RL, DRL, and MARL under identical conditions.
    
    \item \textbf{Narrow Attack Coverage:} Most trust-based frameworks evaluate against one or two attack types. Sophisticated attacks such as Adaptive Adversarial Attacks (learning-based adversaries) and Time-Delayed Poisoning (sleeper agents) remain largely unexplored.
    
    \item \textbf{Coordinated Attack Defense:} While MARL has been studied for multi-agent coordination, its application to defending against \textit{coordinated} adversarial attacks on blockchain trust systems lacks empirical evaluation.
    
    \item \textbf{Temporal Attack Vulnerabilities:} Time-delayed attacks where adversaries build trust before attacking represent a realistic threat model that existing frameworks do not adequately address.
\end{enumerate}

This work addresses these gaps by providing a comprehensive evaluation framework comparing three learning paradigms against five attack families, with particular attention to coordinated (CRA) and temporal (TDP) threats that exploit fundamental assumptions in trust-based systems.

\section{System Model and Problem Formulation}
\label{sec:methodology}

We consider a blockchain-enabled Internet of Things (IoT) network composed of multiple nodes, each representing an IoT device. The network employs a permissioned blockchain utilizing a trust-based delegated consensus mechanism. Smart contracts automate trust evaluations and access control policies. To enhance privacy and security, Attribute-Based Access Control (ABAC) integrates with Fully Homomorphic Encryption (FHE), enabling secure computations on encrypted attribute data.

\begin{figure}[htbp]
\centering
\includegraphics[width=\linewidth]{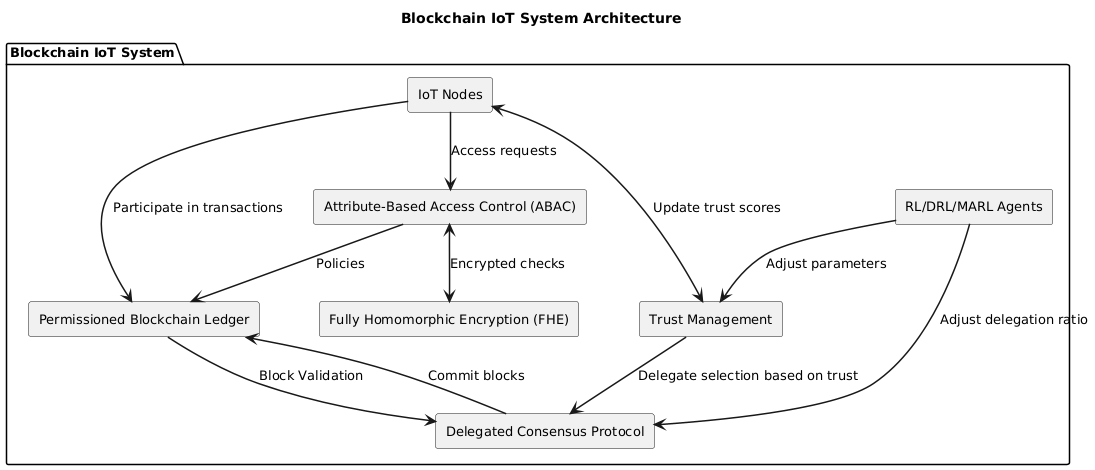}
\caption{Blockchain IoT System Architecture integrating FHE-secured ABAC with trust-based delegated consensus and reinforcement learning defense.}
\label{fig:system_architecture}
\end{figure}

\subsection{Blockchain and Trust Model}

Each node participates in transaction validation and block creation while maintaining dynamic trust scores updated based on observed behaviors. We adopt a Bayesian trust model where each node $i$ is characterized by parameters $(\alpha_i, \beta_i)$ of a Beta distribution \cite{josang2002beta}, yielding a trust score $\tau_i = \alpha_i / (\alpha_i + \beta_i)$. This probabilistic representation captures both the estimated trustworthiness and the uncertainty associated with limited observations.

Trust updates follow Bayesian inference principles. When a node exhibits valid behavior, we increment $\alpha_i$ by $\delta_{\text{valid}}$, increasing its expected trust. Invalid behavior increments $\beta_i$ by $\delta_{\text{invalid}}$, while confirmed malicious behavior triggers both a $\beta_i$ increment of $\delta_{\text{malicious}}$ and a multiplicative decay $\alpha_i \leftarrow \gamma \cdot \alpha_i$ where $\gamma < 1$. This asymmetric update scheme reflects the security principle that trust should be harder to build than to lose.

Trust scores directly influence delegate selection for consensus. We employ Thompson Sampling, where for each selection round we draw samples \cite{thompson1933likelihood} from each node's Beta distribution and select the top-$k$ nodes as delegates. The parameter $k$ is determined by the current delegation ratio, which the learning agent dynamically adjusts. This approach naturally balances exploitation of established high-trust nodes with exploration of nodes whose trustworthiness remains uncertain.

\subsection{Attribute-Based Access Control with FHE}

The ABAC system restricts transaction processing based on attributes including device roles, permissions, security clearances, and real-time trust scores. Traditional ABAC implementations expose attribute values during policy evaluation, creating potential attack vectors. Our framework addresses this vulnerability by evaluating access control policies using Fully Homomorphic Encryption, expressed as $\text{Decision} = \text{FHE.Decrypt}(\text{FHE.Eval}(f_{\text{policy}}, \text{FHE.Encrypt}(\text{attributes})))$, where $f_{\text{policy}}$ represents the access control policy function.

This encryption scheme ensures that sensitive attribute information remains confidential throughout policy evaluation. Even if an adversary intercepts the encrypted policy evaluation, they cannot determine which attributes triggered acceptance or rejection, preventing targeted attacks that exploit knowledge of specific attribute thresholds or combinations.

\subsection{Network and Node Assumptions}

We assume a heterogeneous IoT network comprising $N = 16$ nodes with varying computational capabilities, representing a typical small-to-medium scale IoT deployment. The malicious ratio $\rho = 0.30$ yields approximately 5 malicious nodes, consistent with Byzantine fault tolerance assumptions that tolerate up to one-third adversarial participants. A trust threshold $\theta = 0.45$ determines malicious classification, set slightly below the neutral value of 0.5 to reduce false negatives at the cost of potential false positives. Communication links are generally reliable for primary consensus activities, though we acknowledge intermittent disruptions typical in IoT environments.

\subsection{Threat Model}
\label{subsec:threat_model}

We consider five distinct attack families targeting trust-based consensus mechanisms, spanning naive, coordinated, adaptive, Byzantine, and temporal threat models. Each attack exploits different assumptions underlying trust-based systems, providing comprehensive coverage of realistic adversarial strategies.

\subsubsection{Naive Malicious Attack (NMA)}

The Naive Malicious Attack represents unsophisticated but persistent threats commonly observed in real-world IoT deployments. Independent malicious nodes randomly disrupt trust scores without explicit coordination, each operating autonomously with probability $p_{\text{attack}}$ of injecting noise into trust evaluations. Malicious nodes randomly penalize honest nodes within their communication range but lack awareness of other attackers' actions. Despite its simplicity, NMA poses a baseline threat that any robust system must handle, as it requires no adversarial infrastructure or communication channels.

\subsubsection{Collusive Rumor Attack (CRA)}

The Collusive Rumor Attack exploits the fundamental assumption that trust endorsements are independent observations. Malicious nodes coordinate to inflate each other's trust scores through mutual endorsement: $\tau_m \leftarrow \tau_m + \sum_{m' \in \mathcal{M}, m' \neq m} \delta_{\text{boost}}$ for all $m \in \mathcal{M}$, where $\mathcal{M}$ denotes the set of malicious nodes. Simultaneously, they target high-trust honest nodes for coordinated penalization: $\tau_h \leftarrow \tau_h - |\mathcal{M}| \cdot \delta_{\text{penalty}}$ for selected $h \in \mathcal{H}$. This dual strategy aims to elevate malicious nodes into delegate positions while excluding legitimate participants, ultimately compromising consensus integrity.

\subsubsection{Adaptive Adversarial Attack (AAA)}

Inspired by adversarial machine learning literature \cite{standen2025adversarial}, the Adaptive Adversarial Attack implements intelligent attackers that learn defense patterns and dynamically adjust strategies. Unlike static attacks, AAA maintains performance metrics for multiple attack strategies and selects among them using an $\epsilon$-greedy policy: with probability $\epsilon$ it explores random strategies, otherwise it exploits the historically most successful approach.

The attack rotates among five strategies. Gradient exploitation estimates trust update patterns and exploits predictable dynamics. Slow poisoning applies imperceptible trust decay to honest nodes, evading threshold-based detection. Strategic cooperation coordinates mutual trust boosting among malicious nodes. Mimicry copies behavioral patterns of the highest-trust honest nodes to appear legitimate. Temporal coordination synchronizes attacks at strategic intervals to maximize impact while minimizing detection windows. This multi-strategy approach makes AAA particularly challenging for defenses that adapt to specific attack patterns.

\subsubsection{Byzantine Fault Injection (BFI)}

Byzantine Fault Injection implements classical Byzantine behavior \cite{clement_making_2009} enhanced with modern attack techniques. The core Byzantine strategy involves equivocation—sending conflicting messages to different nodes to split consensus and prevent agreement. BFI amplifies this through Sybil techniques, where each Byzantine node maintains $k$ virtual identities that appear as independent participants, multiplying adversarial influence by factor $k$.

Additionally, BFI employs eclipse attacks that isolate specific honest nodes by controlling their network view, feeding them false information about network state. Coordinated strikes synchronize all Byzantine nodes for periodic high-impact attacks. The attack adaptively phases between AGGRESSIVE mode when Byzantine trust is high, STRATEGIC mode for balanced operation, and RECOVERY mode when detection risk increases, demonstrating sophisticated threat adaptation.

\subsubsection{Time-Delayed Poisoning (TDP)}

Time-Delayed Poisoning implements a ``sleeper agent'' attack pattern representing Advanced Persistent Threats (APT). Unlike immediate attacks, TDP operates in two distinct phases. During the dormant phase (episodes $1$ to $T_{\text{activate}} = 25$), sleeper agents behave as model citizens: $\tau_m \leftarrow \tau_m + \delta_{\text{valid}}$, accumulating trust through consistent valid behavior. The system perceives these nodes as highly reliable, potentially selecting them as delegates.

Upon activation in phase two, sleeper agents leverage their accumulated trust and delegate positions to launch coordinated attacks: $\tau_h \leftarrow \tau_h - |\mathcal{M}| \cdot \delta_{\text{attack}}$ targeting honest nodes. TDP exploits systems that weight recent behavior heavily and lack long-term memory of trust patterns. The attack is particularly insidious because the very mechanisms designed to reward good behavior—trust accumulation and delegate selection—become vectors for compromise.

\begin{figure}[htbp]
\centering
\includegraphics[width=\linewidth]{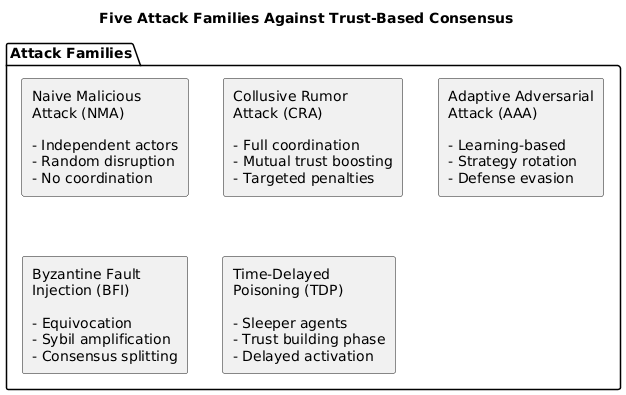}
\caption{Five Attack Families: NMA operates independently, CRA coordinates trust manipulation, AAA adapts to defenses, BFI splits consensus, and TDP exploits temporal trust dynamics.}
\label{fig:attack_scenarios}
\end{figure}

Table~\ref{tab:attack_comparison} summarizes the distinguishing characteristics of each attack family, highlighting their coordination requirements, adaptive capabilities, temporal dynamics, and primary targets.

\begin{table}[htbp]
\centering
\caption{Comparison of Attack Families}
\label{tab:attack_comparison}
\begin{tabular}{|l|c|c|c|c|}
\hline
\textbf{Attack} & \textbf{Coordination} & \textbf{Adaptive} & \textbf{Temporal} & \textbf{Primary Target} \\
\hline
NMA & None & No & No & Random nodes \\
CRA & Full & No & No & High-trust honest nodes \\
AAA & Partial & Yes & No & Defense mechanism \\
BFI & Full & Partial & No & Consensus agreement \\
TDP & Full & No & Yes & Trust memory \\
\hline
\end{tabular}
\end{table}

\subsection{Problem Formulation}

The primary objective is to develop a learning-based adaptive defense mechanism that dynamically adjusts delegation ratios to mitigate the impact of all five attack families. We formalize this as a Markov Decision Process (MDP) defined by the tuple $(\mathcal{S}, \mathcal{A}, \mathcal{P}, \mathcal{R}, \gamma)$.

\subsubsection{State Space}

The state vector $s_t \in \mathbb{R}^{16}$ captures comprehensive network and trust information at each decision step. The first seven components characterize the trust distribution: mean trust $\bar{\tau}$, variance $\sigma^2_{\tau}$, skewness, median $\tilde{\tau}$, range, interquartile range, and coefficient of variation. These statistics enable the agent to detect anomalies such as bimodal distributions (indicating successful honest/malicious separation) or compressed ranges (suggesting collusion success).

The remaining components capture operational metrics: normalized verified transactions and blockchain length measure system throughput, the honest-to-malicious ratio provides ground truth context during training, fractions of low-trust and high-trust nodes indicate detection progress, delegation efficiency reflects current policy, transaction throughput rate measures performance, recent block creation rate captures temporal dynamics, and a collusion detection score specifically flags coordinated attack signatures. This 16-dimensional representation provides sufficient information for the agent to distinguish between attack types and adapt its defense strategy accordingly.

\subsubsection{Action Space}

The discrete action space $\mathcal{A} = \{a_0, a_1, a_2\}$ controls delegation ratio adjustment through multiplicative factors. Action $a_0$ decreases the delegation ratio by factor 0.9, reducing the number of delegates and potentially excluding compromised nodes. Action $a_1$ maintains the current ratio, appropriate when the current policy performs well. Action $a_2$ increases the ratio by factor 1.1, expanding the delegate pool to improve throughput when the network appears secure. The delegation ratio is clipped to $[0.1, 1.0]$ to ensure meaningful delegation while preventing degenerate cases.

\subsubsection{Reward Function}

The reward function balances detection accuracy with operational efficiency. The primary component rewards high F1-scores, computed as $w_{\text{F1}} \cdot \text{F1} \cdot 100$ with $w_{\text{F1}} = 0.7$, emphasizing correct classification of both honest and malicious nodes. A secondary component $w_{\text{step}} \cdot R_{\text{step}} / 100$ with $w_{\text{step}} = 0.3$ rewards operational metrics including throughput and consensus success.

Two penalty terms address specific failure modes. The false negative penalty $w_{\text{FN}} \cdot |\text{FN}|$ with $w_{\text{FN}} = 3.0$ strongly discourages missing malicious nodes, reflecting the asymmetric cost of security failures versus false alarms. For CRA specifically, a collusion penalty activates when the collusion score $\kappa_{\text{collusion}}$ exceeds 2.0, computed as $\min(\kappa_{\text{collusion}} \times 2, 20)$. The collusion score itself measures trust separation: $\kappa_{\text{collusion}} = 1 / |\bar{\tau}_{\mathcal{H}} - \bar{\tau}_{\mathcal{M}}|$, approaching infinity when honest and malicious nodes become indistinguishable.

\subsection{Learning Agents}

We comparatively evaluate three reinforcement learning paradigms representing increasing levels of sophistication and coordination capability.

\subsubsection{Tabular Q-Learning (RL)}

Standard Q-learning maintains a tabular value function over discretized states, updating according to the temporal difference rule: $Q(s_t, a_t) \leftarrow Q(s_t, a_t) + \alpha [ r_t + \gamma \max_{a'} Q(s_{t+1}, a') - Q(s_t, a_t) ]$. While simple and interpretable, tabular methods suffer from the curse of dimensionality in our 16-dimensional state space \cite{bellman1957dynamic}, requiring aggressive discretization that may lose critical information. We include RL as a baseline to quantify the value added by deep learning approaches.

\subsubsection{Dueling Double Deep Q-Network (DRL)}

Our DRL implementation uses the Dueling Double DQN (D3QN) architecture \cite{wang2016dueling} \cite{muniswamy2024trust}, which separates value and advantage estimation: $Q(s, a; \theta) = V(s; \theta_V) + ( A(s, a; \theta_A) - \frac{1}{|\mathcal{A}|} \sum_{a'} A(s, a'; \theta_A) )$. The dueling architecture improves learning efficiency by allowing the network to learn state values independent of action effects. Double Q-learning addresses overestimation bias \cite{vanhasselt2016deep} by using separate networks for action selection and evaluation. Experience replay and target network updates provide training stability essential for security-critical applications.

\subsubsection{Multi-Agent Reinforcement Learning (MARL)}

Our MARL framework deploys $N$ independent learning agents, one per network node, each maintaining its own Q-network but observing a shared global state and reward signal. Agent $i$ updates its value function as: $Q_i(s_t, a_t) \leftarrow Q_i(s_t, a_t) + \alpha_i [ r_t + \gamma \max_{a'} Q_i(s_{t+1}, a') - Q_i(s_t, a_t) ]$. This decentralized approach enables emergent coordination where different agents may specialize in detecting different attack patterns. We hypothesize that MARL's distributed learning provides advantages against coordinated attacks (CRA, BFI) where single-agent methods may be overwhelmed.

\begin{figure}[htbp]
\centering
\includegraphics[width=\linewidth]{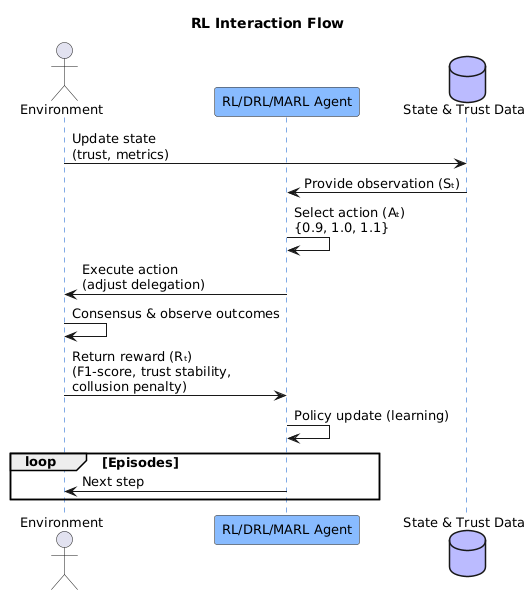}
\caption{Reinforcement Learning Interaction: The agent observes network state, selects delegation ratio adjustment, and receives reward based on detection performance and operational metrics.}
\label{fig:rl_interaction}
\end{figure}

Table~\ref{tab:agent_comparison} contrasts the three paradigms across key dimensions. RL offers simplicity but limited scalability; DRL provides powerful function approximation for complex state spaces; MARL adds coordination capability at the cost of increased training complexity.

\begin{table}[htbp]
\centering
\caption{Comparison of Learning Agents}
\label{tab:agent_comparison}
\begin{tabular}{|l|c|c|c|}
\hline
\textbf{Property} & \textbf{RL} & \textbf{DRL} & \textbf{MARL} \\
\hline
State Representation & Discretized & Continuous & Continuous \\
Function Approximation & Table & Neural Network & Neural Network \\
Number of Agents & 1 & 1 & $N$ (per node) \\
Coordination Capability & None & None & Emergent \\
Scalability & Limited & Good & Good \\
\hline
\end{tabular}
\end{table}

\section{Proposed Approach}
\label{sec:approach}

Our proposed framework integrates three synergistic components to create a comprehensive defense against trust manipulation attacks in blockchain IoT environments. The foundation consists of a trust-based delegated consensus mechanism that dynamically evaluates node reliability. This is secured by Attribute-Based Access Control (ABAC) implemented with Fully Homomorphic Encryption (FHE) to protect sensitive policy evaluations. Finally, reinforcement learning agents—spanning tabular, deep, and multi-agent paradigms—provide adaptive defense that learns optimal responses to the five attack families described in Section~\ref{subsec:threat_model}.

\subsection{Trust-Based Consensus Blockchain}

We utilize a permissioned blockchain with a delegated consensus mechanism combining elements of Delegated Proof-of-Stake and Byzantine Fault Tolerance. Unlike traditional consensus where all nodes participate equally, our approach dynamically selects a consensus committee based on accumulated trust, reducing computational overhead while maintaining security guarantees.

Each node maintains a trust profile characterized by Bayesian parameters $(\alpha, \beta)$ representing positive and negative evidence respectively. The trust score $\tau = \alpha / (\alpha + \beta)$ provides a point estimate, while the full distribution captures uncertainty—critical for distinguishing genuinely trustworthy nodes from those with insufficient observation history. Trust updates occur incrementally following each consensus round: nodes that correctly validate transactions and adhere to protocol receive positive evidence ($\alpha$ increments), while detected misbehavior triggers negative evidence ($\beta$ increments) and, for confirmed malicious activity, multiplicative $\alpha$ decay.

\begin{figure}[htbp]
    \centering
    \includegraphics[width=\linewidth]{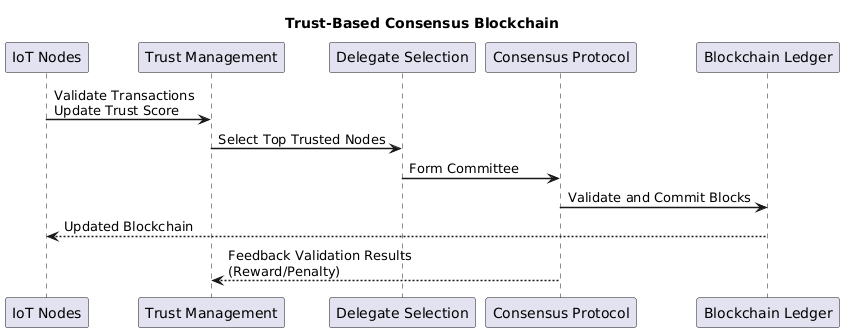}
    \caption{Trust-Based Consensus Blockchain: Nodes accumulate trust through valid behavior, with top-ranked nodes forming the consensus committee. Trust scores are updated based on validation outcomes and protocol adherence.}
    \label{fig:trust}
\end{figure}

Delegate selection employs Thompson Sampling rather than simple top-$k$ selection. For each selection round, we draw samples from each node's Beta distribution and select nodes with the highest sampled values. This approach naturally balances exploitation of established high-trust nodes with exploration of uncertain candidates, preventing adversaries from permanently excluding honest nodes through initial trust suppression. The delegation ratio—the fraction of nodes selected as delegates—is dynamically adjusted by the learning agent based on observed attack patterns. This process is illustrated in Fig.~\ref{fig:trust}.

\subsection{Attribute-Based Access Control with Fully Homomorphic Encryption}

The ABAC mechanism enforces granular access control policies based on device attributes including roles, permissions, security clearances, and real-time trust scores. Traditional ABAC implementations evaluate policies in plaintext, exposing attribute values and policy logic to potential adversaries. Our framework addresses this vulnerability through integration with Fully Homomorphic Encryption, enabling the complete policy evaluation lifecycle to occur on encrypted data.

When a node requests transaction processing, its attributes are encrypted using the FHE public key. The encrypted attributes are then evaluated against encrypted policy conditions using homomorphic operations that preserve the logical structure of the access decision. The result—an encrypted accept/reject decision—is decrypted only by authorized policy enforcement points. This design ensures that even if an adversary compromises intermediate computation nodes, they cannot determine which specific attributes triggered acceptance or rejection, preventing targeted attacks that exploit knowledge of policy thresholds.

\begin{figure}[htbp]
    \centering
    \includegraphics[width=\linewidth]{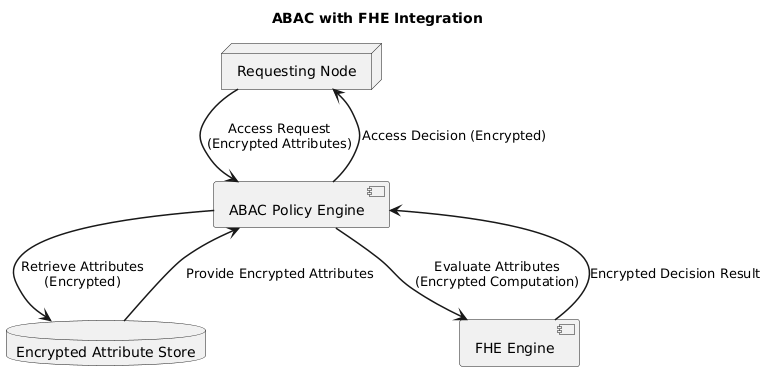}
    \caption{ABAC integrated with Fully Homomorphic Encryption: Access requests are evaluated entirely on encrypted data, preventing attribute leakage even during policy evaluation.}
    \label{fig:ABACFHE}
\end{figure}

The integration proves particularly valuable against Adaptive Adversarial Attacks (AAA), where intelligent adversaries attempt to learn and exploit defense mechanisms. By hiding both attributes and policy logic, FHE-secured ABAC prevents adversaries from inferring the decision boundary through repeated probing. The integration of ABAC and FHE is depicted in Fig.~\ref{fig:ABACFHE}.

\subsection{Reinforcement Learning Defense Framework}

The core innovation of our approach lies in using reinforcement learning to dynamically adapt defense parameters in response to evolving attack patterns. Rather than relying on fixed detection thresholds or static trust update rules, our learning agents observe network state and adjust the delegation ratio to optimize detection performance across all five attack families. We implement and compare three learning paradigms of increasing sophistication.

\subsubsection{Tabular Q-Learning (RL Baseline)}

A classical Q-learning agent serves as our lightweight baseline for delegation ratio control. At every time step, the agent discretizes the 16-dimensional state vector into a finite number of bins and consults a tabular Q-table mapping state-action pairs to expected cumulative rewards. Action selection follows an $\varepsilon$-greedy policy: with probability $\varepsilon$ the agent explores by selecting a random action from $\{0.9, 1.0, 1.1\}$, otherwise it exploits by selecting the action with highest Q-value.

\begin{figure}[htbp]
  \centering
  \includegraphics[width=\linewidth]{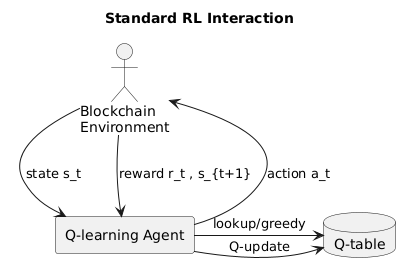}
  \caption{Standard RL interaction flow: The agent discretizes continuous state observations, consults a Q-table for action selection, and updates values via temporal difference learning.}
  \label{fig:rl_agent_flow}
\end{figure}

After the environment returns the reward, the Q-table is updated using temporal difference learning with learning rate $\alpha$ \cite{sutton1988learning} and discount factor $\gamma$. The exploration rate $\varepsilon$ decays over episodes to shift from exploration toward exploitation as the agent accumulates experience. Although simple and interpretable, this agent faces fundamental limitations: aggressive state discretization loses information critical for distinguishing sophisticated attacks, and the tabular representation scales poorly with state dimensionality. We include this baseline to quantify the value added by deep learning approaches and to identify attack types where simple methods suffice. The full interaction cycle is summarized in Fig.~\ref{fig:rl_agent_flow}.

\subsubsection{Dueling Double Deep Q-Network (DRL)}

Our primary defense agent employs the Dueling Double Deep Q-Network (D3QN) architecture, which addresses several limitations of both tabular methods and vanilla deep Q-learning. The architecture separates value estimation into two streams: a value function $V(s)$ estimating the expected return from state $s$ regardless of action, and an advantage function $A(s,a)$ estimating the relative benefit of each action. These streams combine as:
\begin{equation}
    Q(s, a) = V(s) + \left( A(s, a) - \frac{1}{|\mathcal{A}|} \sum_{a'} A(s, a') \right)
\end{equation}

This decomposition improves learning efficiency by allowing the network to learn state values without requiring action-specific feedback, particularly beneficial in states where the choice of action has minimal impact.

The Double Q-learning component addresses overestimation bias inherent in standard Q-learning by using separate networks for action selection and evaluation. The online network selects the best action: $a^* = \arg\max_a Q(s', a; \theta)$, while the target network evaluates it: $Q(s', a^*; \theta^-)$. Target network parameters $\theta^-$ are updated periodically from online parameters $\theta$, providing stable learning targets.

\begin{figure}[htbp]
    \centering
    \includegraphics[width=\linewidth]{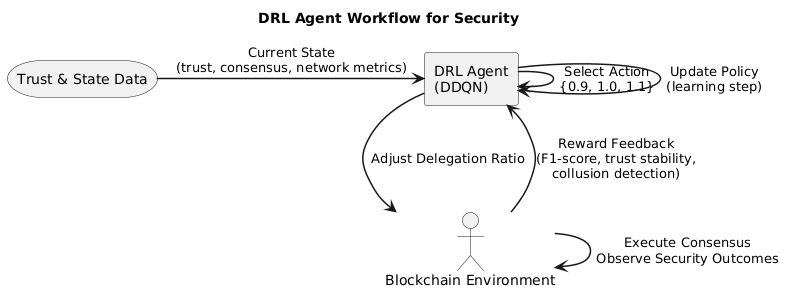}
    \caption{Deep Reinforcement Learning interaction flow: The D3QN agent processes continuous state observations through neural networks, enabling sophisticated pattern recognition for attack detection.}
    \label{fig:drl_agent}
\end{figure}

Experience replay further stabilizes training \cite{mnih2015human} by storing transitions $(s, a, r, s')$ in a buffer and sampling random mini-batches for updates, breaking temporal correlations that can destabilize neural network training. The DRL agent learns to recognize attack signatures in the continuous state space and responds with appropriate delegation adjustments. This dynamic process is illustrated in Fig.~\ref{fig:drl_agent}.

\subsubsection{Multi-Agent Reinforcement Learning (MARL)}

To exploit the inherently decentralized nature of blockchain networks, we deploy a Multi-Agent Reinforcement Learning framework where each validator node runs an independent learning agent. Unlike fully independent learners that may converge to conflicting policies, our agents share neural network parameters while maintaining separate experience buffers, enabling cooperative learning without requiring explicit communication protocols.

Each agent observes the global network state—including trust distributions and consensus metrics—and receives a shared reward reflecting overall security and consensus success. Parameter sharing accelerates convergence by allowing agents to learn from collective experience: a successful defense strategy discovered by one agent immediately benefits all others through shared weights. This approach proves particularly effective against coordinated attacks (CRA, BFI) where single-agent methods may be overwhelmed by synchronized adversarial behavior.

\begin{figure}[htbp]
  \centering
  \includegraphics[width=\linewidth]{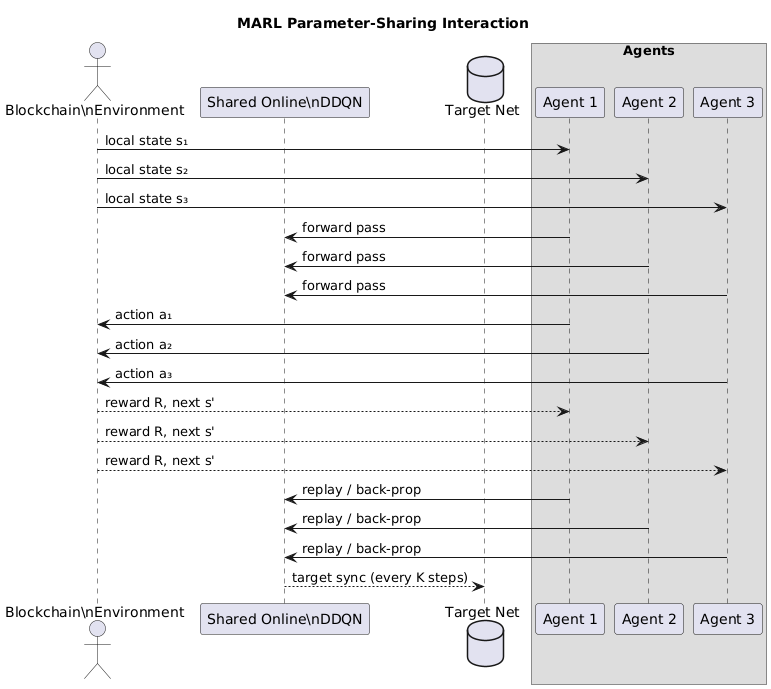}
  \caption{Parameter-sharing MARL \cite{gupta2017cooperative} workflow: Independent agents at each node share neural network parameters, enabling cooperative defense learning against coordinated attacks.}
  \label{fig:marl_agent_flow}
\end{figure}

The distributed learning architecture also provides robustness to partial observation failures. If some agents receive corrupted or incomplete state information—as might occur during Eclipse attacks in BFI—other agents can compensate through the shared policy. Asynchronous updates prevent bottlenecks while periodic synchronization ensures policy coherence. Figure~\ref{fig:marl_agent_flow} illustrates the distributed learning loop and synchronization mechanism.

\subsection{Defense Against Five Attack Families}

The combination of trust-based consensus, FHE-secured ABAC, and adaptive learning provides layered defense against all five attack families. Against Naive Malicious Attacks (NMA), the Bayesian trust model naturally down-weights unreliable nodes through accumulated negative evidence, while Thompson Sampling exploration prevents permanent exclusion of temporarily penalized honest nodes.

Collusive Rumor Attacks (CRA) are countered through the collusion detection score embedded in the state representation, which flags anomalously small trust gaps between known-honest and known-malicious nodes. The MARL framework proves particularly effective here, as distributed agents can detect coordinated trust inflation patterns that single observers might miss.

Adaptive Adversarial Attacks (AAA) face the combined challenge of FHE-hidden policy logic and continuously adapting defense policies. While AAA attempts to learn defense patterns, the RL agents simultaneously learn counter-strategies, creating an adversarial game where our results show DRL and MARL maintain advantage.

Byzantine Fault Injection (BFI) attacks are mitigated through the trust mechanism's integration with consensus: equivocating nodes accumulate negative evidence from conflicting reports, while Sybil identities inherit the trust limitations of their parent Byzantine node. The delegation mechanism excludes low-trust nodes from consensus committees, limiting Byzantine influence.

Time-Delayed Poisoning (TDP) represents the most challenging threat, as sleeper agents accumulate genuine positive evidence during their dormant phase. Our results demonstrate that no learning approach fully counters activated TDP attacks, highlighting an important limitation and direction for future work involving longer-term trust memory mechanisms.

Table~\ref{tab:defense_mapping} summarizes how each framework component contributes to defending against specific attack families.

\begin{table}[htbp]
\centering
\caption{Defense Component Effectiveness Against Attack Families}
\label{tab:defense_mapping}
\begin{tabular}{|l|c|c|c|c|c|}
\hline
\textbf{Component} & \textbf{NMA} & \textbf{CRA} & \textbf{AAA} & \textbf{BFI} & \textbf{TDP} \\
\hline
Bayesian Trust & \checkmark & \checkmark & \checkmark & \checkmark & Limited \\
Thompson Sampling & \checkmark & -- & -- & -- & -- \\
FHE-ABAC & -- & -- & \checkmark & -- & -- \\
Collusion Detection & -- & \checkmark & -- & -- & -- \\
DRL Adaptation & \checkmark & \checkmark & \checkmark & \checkmark & Limited \\
MARL Coordination & -- & \checkmark & \checkmark & \checkmark & Limited \\
\hline
\end{tabular}
\end{table}

\section{Attack Models}
\label{sec:attacks}

We analyze five adversarial scenarios targeting trust-based blockchain consensus mechanisms. These attacks span a spectrum from unsophisticated independent disruptions to highly coordinated temporal strategies, providing comprehensive coverage of realistic threat models in blockchain IoT environments.

\subsection{Naive Malicious Attack (NMA)}

The Naive Malicious Attack represents the simplest adversarial threat model, where malicious nodes independently perform random disruptions without explicit coordination. Each malicious node operates autonomously, randomly perturbing trust evaluations through incorrect validations and false feedback. The attack requires no communication infrastructure between adversaries and no knowledge of system defense mechanisms.

During each consensus round, a malicious node with probability $p_{\text{attack}}$ injects noise into its trust evaluations, randomly penalizing honest nodes within its observation range. The uncoordinated nature means that malicious activities may occasionally cancel out—one attacker might penalize a node that another inadvertently boosted. Despite this inefficiency, NMA poses a persistent baseline threat that any robust system must handle, as it requires minimal adversarial capability and naturally emerges in compromised IoT deployments where devices are independently infected.

The primary impact of NMA is degraded system stability rather than systematic compromise. Trust scores exhibit increased variance, and consensus may occasionally include unreliable delegates, but the lack of coordination prevents adversaries from achieving sustained control over the delegation process. Detection is relatively straightforward: nodes consistently providing incorrect validations accumulate negative trust evidence and are eventually excluded from delegation.

\begin{figure}[htbp]
\centering
\includegraphics[width=\linewidth]{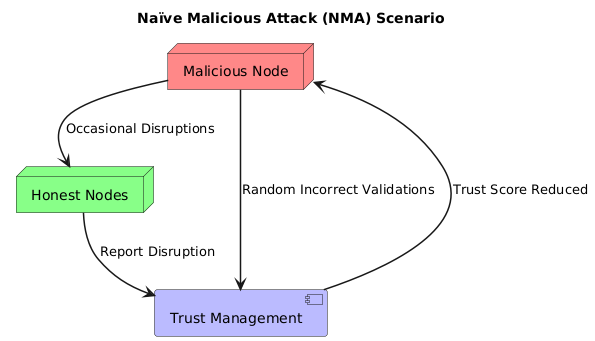}
\caption{Naive Malicious Attack (NMA): Independent malicious nodes randomly disrupt trust evaluations without coordination, causing intermittent system instability.}
\label{fig:nma_attack}
\end{figure}

\subsection{Collusive Rumor Attack (CRA)}

The Collusive Rumor Attack elevates the threat model through explicit coordination among malicious nodes. Rather than acting independently, CRA adversaries form a coalition that strategically manipulates the trust system through two complementary mechanisms: mutual trust inflation and targeted honest node suppression.

In the inflation phase, each malicious node $m \in \mathcal{M}$ provides falsified positive validations for other coalition members, artificially boosting their trust scores according to $\tau_m \leftarrow \tau_m + \sum_{m' \in \mathcal{M}, m' \neq m} \delta_{\text{boost}}$. Since trust systems typically assume independent observations, this coordinated endorsement appears as multiple independent confirmations of trustworthiness. Simultaneously, the coalition identifies high-trust honest nodes—those most likely to be selected as delegates—and coordinates negative feedback: $\tau_h \leftarrow \tau_h - |\mathcal{M}| \cdot \delta_{\text{penalty}}$ for targeted $h \in \mathcal{H}$.

The strategic goal is delegate capture: by elevating malicious trust while suppressing honest competitors, CRA aims to achieve majority representation in consensus committees. Once achieved, the coalition can manipulate transaction validation, censor specific transactions, or introduce invalid blocks. The attack exploits the fundamental assumption underlying reputation systems—that trust endorsements reflect independent observations of actual behavior.

Detection requires identifying statistical anomalies in trust update patterns. Our framework incorporates a collusion score $\kappa = 1/|\bar{\tau}_{\mathcal{H}} - \bar{\tau}_{\mathcal{M}}|$ that flags suspiciously small gaps between honest and malicious trust distributions, triggering defensive responses from the learning agent.

\begin{figure}[htbp]
\centering
\includegraphics[width=0.5\linewidth]{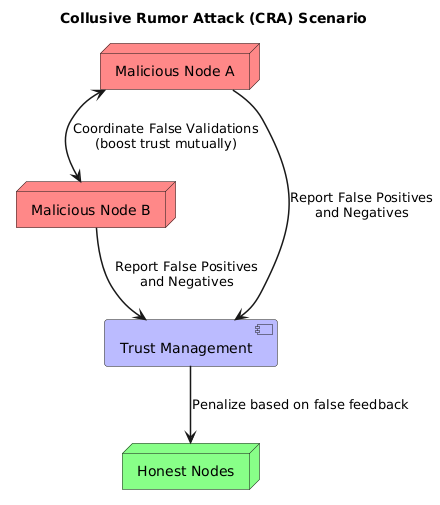}
\caption{Collusive Rumor Attack (CRA): Coordinated malicious nodes mutually inflate trust scores while systematically penalizing honest nodes to capture delegate positions.}
\label{fig:cra_attack}
\end{figure}

\subsection{Adaptive Adversarial Attack (AAA)}

The Adaptive Adversarial Attack introduces intelligence into the adversarial model, implementing attackers that learn defense patterns and dynamically adjust their strategies. Inspired by adversarial machine learning research \cite{standen2025adversarial}, AAA represents sophisticated adversaries capable of observing defense responses and optimizing their attack approach accordingly.

AAA maintains a portfolio of five attack strategies and selects among them using reinforcement learning principles. Gradient exploitation estimates trust update dynamics and manipulates behavior to maximize trust gain while minimizing detection probability. Slow poisoning applies imperceptible trust decay to honest nodes over extended periods, evading threshold-based detection by keeping individual perturbations below observable limits. Strategic cooperation coordinates trust boosting among malicious nodes, similar to CRA but with adaptive intensity based on detection risk. Mimicry observes the behavioral patterns of highest-trust honest nodes and replicates them, making malicious nodes statistically indistinguishable from legitimate participants. Temporal coordination synchronizes attack activities to specific time windows, maximizing impact while minimizing exposure duration.

Strategy selection follows an $\varepsilon$-greedy policy with performance tracking. The attack maintains success metrics for each strategy and preferentially selects historically effective approaches while maintaining exploration probability $\varepsilon$ to discover new vulnerabilities. The attack phases through exploration (learning defense patterns), exploitation (applying learned strategies), and evasion (avoiding detected patterns) as the engagement progresses.

AAA poses particular challenges because it creates an adversarial game: as our defense agents learn to counter specific attack patterns, AAA learns to avoid those patterns. The equilibrium outcome depends on relative learning rates and the information asymmetry between attacker and defender. Our FHE-secured ABAC provides crucial advantage by hiding policy logic from adversarial observation.

\begin{figure}[htbp]
\centering
\includegraphics[width=\linewidth]{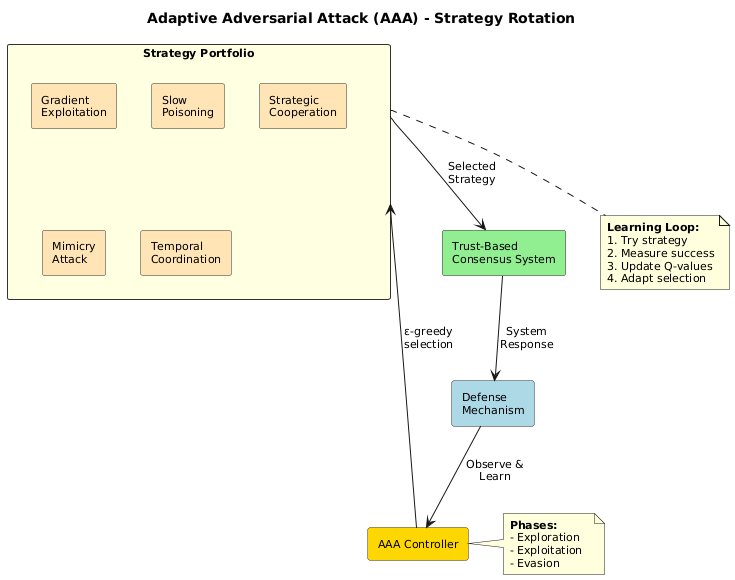}
\caption{Adaptive Adversarial Attack (AAA): Intelligent attackers learn defense patterns and rotate among five strategies—gradient exploitation, slow poisoning, strategic cooperation, mimicry, and temporal coordination—to evade detection.}
\label{fig:aaa_attack}
\end{figure}

\subsection{Byzantine Fault Injection (BFI)}

Byzantine Fault Injection implements classical Byzantine behavior \cite{lamport1982byzantine} enhanced with modern amplification techniques. The attack derives from the Byzantine Generals Problem \cite{clement_making_2009}, where faulty nodes may exhibit arbitrary behavior including sending conflicting information to different observers. BFI operationalizes this theoretical model with practical attack mechanisms designed to maximize consensus disruption.

The core Byzantine strategy is equivocation: sending conflicting votes or validations to different network participants. When node $A$ receives a "valid" vote from Byzantine node $B$ while node $C$ receives an "invalid" vote for the same transaction, the network cannot reach consistent agreement. Repeated equivocation fragments network state, potentially creating forks or stalling consensus entirely. Our framework mitigates equivocation through trust penalties when conflicting messages are detected, but sophisticated timing can delay detection.

BFI amplifies Byzantine influence through Sybil techniques, where each physical Byzantine node maintains $k$ virtual identities that appear as independent network participants. If the system's Byzantine tolerance threshold assumes $f$ faulty nodes among $n$ total, Sybil amplification effectively multiplies adversarial presence by factor $k$, potentially exceeding tolerance bounds. Our trust mechanism partially counters Sybil attacks by requiring each identity to independently accumulate trust through validated behavior, but the amplified initial presence provides coordination advantages.

Eclipse attacks complement equivocation by isolating specific honest nodes from accurate network state. Byzantine nodes controlling an honest node's network connections feed it false information about trust scores, transaction status, and consensus progress. The eclipsed node may make decisions based on fabricated reality, potentially validating invalid transactions or rejecting legitimate ones. Detection requires cross-referencing information through multiple independent channels.

The attack adaptively phases between aggressive, strategic, and recovery modes based on current Byzantine trust levels, demonstrating tactical sophistication beyond simple fault injection.

\begin{figure}[htbp]
\centering
\includegraphics[width=\linewidth]{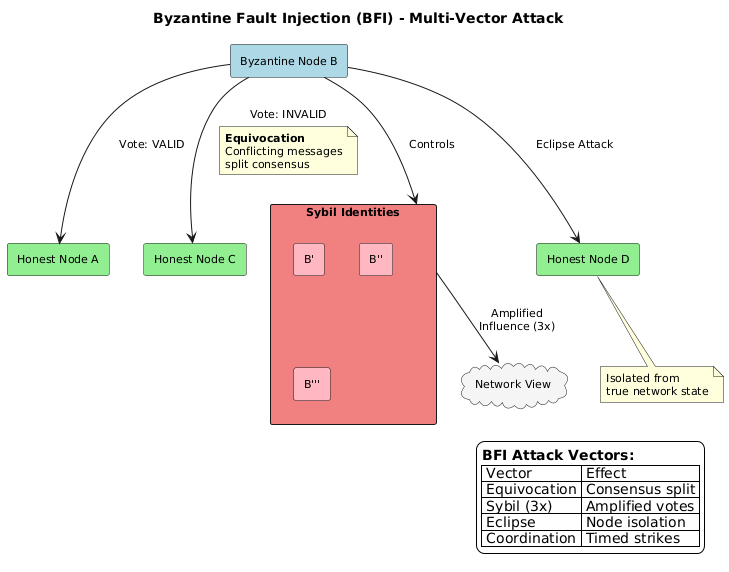}
\caption{Byzantine Fault Injection (BFI): Byzantine nodes employ equivocation, Sybil amplification, and eclipse attacks to split consensus and fragment network state.}
\label{fig:bfi_attack}
\end{figure}

\subsection{Time-Delayed Poisoning (TDP)}

Time-Delayed Poisoning implements the most insidious attack pattern: sleeper agents that invest in building genuine trust before activating coordinated attacks. TDP models Advanced Persistent Threats (APT) \cite{chen2014study} where adversaries prioritize long-term strategic positioning over immediate disruption, exploiting the temporal assumptions underlying trust accumulation.

The attack operates in two distinct phases. During the dormant phase spanning episodes 1 through $T_{\text{activate}}$ (set to 25 in our experiments), sleeper agents behave as model network citizens. They correctly validate transactions, adhere to protocols, and accumulate positive trust evidence indistinguishable from honest behavior: $\tau_m \leftarrow \tau_m + \delta_{\text{valid}}$. The trust system, designed to reward consistent good behavior, elevates these nodes to high-trust status. They may be selected as delegates, granted elevated permissions, and excluded from suspicion lists.

Upon activation in phase two, sleeper agents leverage their accumulated trust and privileged positions to launch coordinated attacks. High-trust attackers selected as delegates can directly manipulate consensus. Even if not delegates, their trusted status means negative feedback they provide against honest nodes carries disproportionate weight. The coordinated attack targets honest nodes for trust suppression: $\tau_h \leftarrow \tau_h - |\mathcal{M}| \cdot \delta_{\text{attack}}$, while sleeper agents continue mutual support.

TDP exploits a fundamental tension in trust system design. Systems must weight recent behavior to enable recovery from past mistakes and adaptation to changing conditions. However, this recency bias creates vulnerability to adversaries willing to invest in extended deception periods. The attack is particularly effective because the defense mechanisms that reward good behavior—trust accumulation, delegate selection, permission elevation—become the vectors for compromise.

Our experimental results demonstrate that TDP poses catastrophic consequences for all learning approaches once activated, with F1 scores dropping to 0.11–0.16. This finding highlights critical limitations of current trust mechanisms and motivates future research into longer-term trust memory and behavioral consistency analysis.

\begin{figure}[htbp]
\centering
\includegraphics[width=\linewidth]{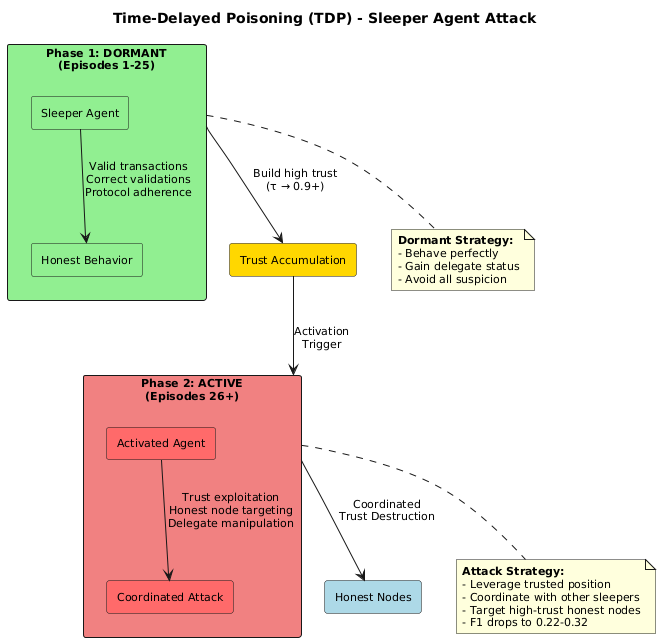}
\caption{Time-Delayed Poisoning (TDP): Sleeper agents behave honestly during the dormant phase to accumulate trust, then activate coordinated attacks from positions of elevated privilege.}
\label{fig:tdp_attack}
\end{figure}

\subsection{Comparative Analysis of Attack Families}

The five attack families span a spectrum of sophistication, coordination, and temporal dynamics, each exploiting different assumptions in trust-based consensus systems. Table~\ref{tab:attack_detailed_comparison} summarizes their distinguishing characteristics.

\begin{table}[htbp]
\centering
\caption{Detailed Comparison of Attack Families}
\label{tab:attack_detailed_comparison}
\begin{tabular}{|l|c|c|c|c|c|}
\hline
\textbf{Characteristic} & \textbf{NMA} & \textbf{CRA} & \textbf{AAA} & \textbf{BFI} & \textbf{TDP} \\
\hline
Coordination Level & None & Full & Partial & Full & Full \\
Adversarial Intelligence & None & Low & High & Medium & Low \\
Temporal Strategy & None & None & Adaptive & None & Phased \\
Primary Target & Random & High-trust honest & Defense mechanism & Consensus & Trust memory \\
Required Infrastructure & Minimal & Communication & Learning capability & Network control & Patience \\
Detection Difficulty & Low & Medium & High & Medium & Very High \\
\hline
\end{tabular}
\end{table}

NMA and CRA differ primarily in coordination: NMA's independent actors cause intermittent disturbances, while CRA's strategic cooperation enables systematic trust manipulation. AAA elevates the threat through adaptive learning, creating an adversarial game against defense mechanisms. BFI targets consensus integrity directly through Byzantine behavior, potentially fragmenting network state regardless of trust scores. TDP exploits temporal assumptions, demonstrating that trust systems optimized for responsiveness become vulnerable to patient adversaries.

The progression from NMA through TDP represents increasing sophistication and decreasing detectability. Simple defenses suffice against NMA, but TDP defeats all evaluated approaches once activated. This spectrum motivates our multi-paradigm learning evaluation: different agents may excel against different threat profiles, and understanding these patterns informs deployment decisions for real-world blockchain IoT systems.

\begin{figure}[htbp]
\centering
\includegraphics[width=\linewidth]{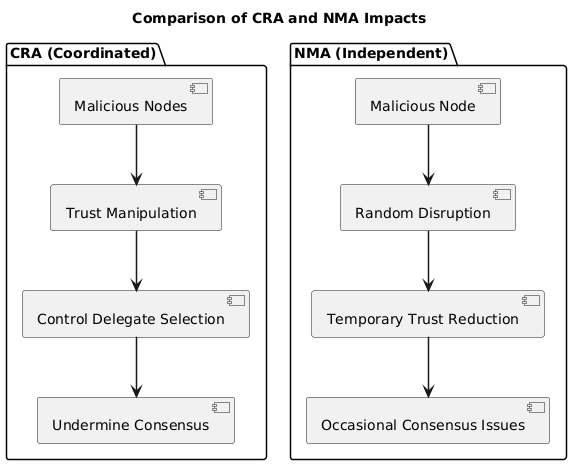}
\caption{Comparative impact analysis showing the spectrum from uncoordinated (NMA) through coordinated (CRA, BFI) to adaptive (AAA) and temporal (TDP) attack strategies.}
\label{fig:attack_comparison}
\end{figure}

\section{Experimental Setup}
\label{sec:experimental}

Our experimental framework systematically evaluates three reinforcement learning paradigms against five attack families, producing 15 distinct agent-attack combinations. We designed the setup to enable fair comparison while maintaining computational tractability.

\subsection{Simulation Environment}

We simulate a permissioned blockchain-enabled IoT network where each node represents an IoT device with heterogeneous capabilities in computing power, storage \cite{atzori2010internet}, and communication reliability. The simulation implements the complete framework described in Section~\ref{sec:approach}, including Bayesian trust management, FHE-secured ABAC policy evaluation, and the Trust-Based Delegated Consensus mechanism.

The implementation uses Python 3.10 with PyTorch 2.0 for neural network components. Experiments were conducted on a system equipped with an NVIDIA Tesla T4 GPU (16GB VRAM), 4-core CPU, and 16GB system RAM. The complete source code, trained model checkpoints, and detailed per-episode results are available as supplementary materials.\footnote{Code and data available at: \url{https://github.com/soham-padia/blockchain-iot-trust}}

\subsection{Network Configuration}

Table~\ref{tab:simulation_params} summarizes the network and simulation parameters. We selected 16 nodes to balance computational tractability with sufficient network complexity for meaningful trust dynamics. The 30\% malicious ratio (approximately 5 malicious nodes) aligns with Byzantine fault tolerance assumptions while presenting a realistic adversarial challenge.

\begin{table}[htbp]
\centering
\caption{Simulation Parameters}
\label{tab:simulation_params}
\begin{tabular}{|l|c|l|}
\hline
\textbf{Parameter} & \textbf{Value} & \textbf{Rationale} \\
\hline
Total Nodes ($N$) & 16 & Balance complexity and tractability \\
Malicious Ratio ($\rho$) & 0.30 & Byzantine tolerance threshold \\
Trust Threshold ($\theta$) & 0.45 & Favor detection over false positives \\
Episodes & 50--100 & Sufficient for convergence \\
Steps per Episode & 100 & Adequate trust evolution time \\
Random Seed & 42 & Reproducibility \\
\hline
\end{tabular}
\end{table}

Trust scores are initialized using a Bayesian Beta distribution with parameters $\alpha_0 = 8.0$ and $\beta_0 = 8.0$, yielding initial trust of approximately 0.5 with added Gaussian noise ($\sigma = 0.12$) to prevent identical starting conditions. This initialization ensures that nodes must demonstrate trustworthy behavior to achieve delegate status rather than inheriting it by default.

\subsection{Learning Agent Configuration}

We evaluate three reinforcement learning paradigms, each configured for fair comparison with equivalent state representations and reward structures.

\subsubsection{Tabular Q-Learning (RL)}

The baseline RL agent discretizes the 16-dimensional continuous state space into bins and maintains a tabular Q-function. We use learning rate $\alpha = 0.1$, discount factor $\gamma = 0.99$, and $\varepsilon$-greedy exploration with initial $\varepsilon = 1.0$ decaying to $\varepsilon_{\min} = 0.05$ over training. State discretization uses 10 bins per dimension for the most informative features (mean trust, variance, collusion score) and 5 bins for others.

\subsubsection{Dueling Double DQN (DRL)}

The DRL agent employs a neural network with architecture: input layer (16 units) $\rightarrow$ hidden layers (128, 64 units with ReLU activation) $\rightarrow$ dueling streams (value: 32$\rightarrow$1, advantage: 32$\rightarrow$3). We use Adam optimizer with learning rate $5 \times 10^{-4}$, experience replay buffer size 10,000, batch size 64, and target network update frequency of 100 steps.

\subsubsection{Multi-Agent RL (MARL)}

The MARL framework deploys 16 agents (one per node) sharing neural network parameters identical to the DRL architecture. Agents observe global state and receive shared rewards, enabling implicit coordination. Experience buffers are maintained independently per agent with synchronized parameter updates every 10 steps.

Table~\ref{tab:agent_hyperparams} summarizes the hyperparameter configurations.

\begin{table}[htbp]
\centering
\caption{Agent Hyperparameters}
\label{tab:agent_hyperparams}
\begin{tabular}{|l|c|c|c|}
\hline
\textbf{Parameter} & \textbf{RL} & \textbf{DRL} & \textbf{MARL} \\
\hline
Learning Rate ($\alpha$) & 0.10 & $5 \times 10^{-4}$ & $5 \times 10^{-4}$ \\
Discount Factor ($\gamma$) & 0.99 & 0.99 & 0.99 \\
Initial Exploration ($\varepsilon_0$) & 1.0 & 1.0 & 1.0 \\
Final Exploration ($\varepsilon_{\min}$) & 0.05 & 0.05 & 0.05 \\
Replay Buffer Size & N/A & 10,000 & 10,000 \\
Batch Size & N/A & 64 & 64 \\
Hidden Layers & N/A & [128, 64] & [128, 64] \\
\hline
\end{tabular}
\end{table}

\subsection{Attack Configurations}

Each attack family is configured with parameters calibrated to produce challenging but not overwhelming adversarial pressure, enabling meaningful comparison of agent capabilities. Table~\ref{tab:attack_configs} details the attack-specific parameters.

\begin{table}[htbp]
\centering
\caption{Attack Configuration Parameters}
\label{tab:attack_configs}
\begin{tabular}{|l|l|c|}
\hline
\textbf{Attack} & \textbf{Key Parameters} & \textbf{Values} \\
\hline
\multirow{2}{*}{NMA} & Attack probability & 0.5 \\
 & Noise level & 0.5 \\
\hline
\multirow{2}{*}{CRA} & Coordination intensity & 0.85 \\
 & Attack frequency & Every 2 steps \\
\hline
\multirow{3}{*}{AAA} & Strategy count & 5 \\
 & Exploration decay ($\varepsilon$) & 0.98 \\
 & Trust manipulation factor & 0.12 \\
\hline
\multirow{3}{*}{BFI} & Equivocation rate & 0.90 \\
 & Sybil amplification & 4$\times$ \\
 & Coordination window & 6 steps \\
\hline
\multirow{3}{*}{TDP} & Activation episode & 25 \\
 & Attack intensity & 0.75 \\
 & Target ratio & 35\% of honest nodes \\
\hline
\end{tabular}
\end{table}

\subsection{Evaluation Protocol}

For each of the 15 agent-attack combinations, we execute complete training runs and evaluate final performance using multiple metrics. To account for stochastic variation, we report results averaged over the final 10 episodes after convergence, with standard deviations where applicable.

\subsubsection{Detection Metrics}

The primary evaluation metric is the \textbf{F1-score} for malicious node detection, computed from the confusion matrix at episode conclusion. A node is classified as malicious if its trust score falls below threshold $\theta = 0.45$. We additionally report precision (fraction of detected nodes that are truly malicious) and recall (fraction of malicious nodes successfully detected). The confusion matrix provides detailed breakdown of true positives, true negatives, false positives, and false negatives.

\subsubsection{Learning Metrics}

\textbf{Cumulative reward} per episode tracks learning progress and convergence behavior. Reward incorporates F1-score, operational metrics, and attack-specific penalties as defined in Section~\ref{sec:methodology}. We analyze reward curves to assess convergence speed, stability, and response to attack activation (particularly relevant for TDP).

\subsubsection{Operational Metrics}

\textbf{Transaction throughput} measures successful transactions processed per episode, reflecting system utility under attack. \textbf{Blockchain length} tracks cumulative block creation as a measure of sustained consensus. \textbf{Trust separation}—the gap between mean honest and mean malicious trust scores—indicates detection margin quality.

\subsection{Experimental Procedure}

Each experiment follows a standardized procedure to ensure comparability:

\begin{enumerate}
    \item Initialize network with 16 nodes, randomly assign 5 as malicious
    \item Initialize trust scores with Bayesian priors plus noise
    \item Reset attack instance to clear any persistent state
    \item Execute training for specified episodes (50 for standard attacks, 100 for TDP to capture post-activation behavior)
    \item Record per-step metrics: trust scores, rewards, F1-scores
    \item Save final confusion matrix and aggregate statistics
    \item Export trained model checkpoint for reproducibility
\end{enumerate}

The complete experimental matrix comprises 15 combinations (3 agents $\times$ 5 attacks), with each combination requiring approximately 10--15 minutes of computation time. Total experimental runtime is approximately 4 hours including result processing and visualization generation.

\subsection{Supplementary Materials}

Given the extensive experimental results (15 combinations $\times$ multiple metrics), we provide comprehensive supplementary materials:

\begin{itemize}
    \item \textbf{Source Code:} Complete Python implementation including all attack models, learning agents, and evaluation scripts
    \item \textbf{Trained Models:} Checkpoint files for all 15 agent-attack combinations
    \item \textbf{Detailed Results:} CSV files containing per-episode metrics for all experiments
    \item \textbf{Extended Figures:} High-resolution versions of all confusion matrices, reward curves, and throughput plots
\end{itemize}

These materials enable full reproducibility and support extended analysis beyond the representative results presented in Section~\ref{sec:results}.
%=================================================================
\section{Results and Analysis}
\label{sec:results}

This section presents experimental results for all 15 agent-attack combinations, analyzing detection performance, learning dynamics, and the relative effectiveness of each defense paradigm. We organize findings around key research questions: (1) How do agents compare across attack types? (2) Which attacks pose the greatest challenge? (3) Does multi-agent learning provide measurable advantages?

\subsection{Detection Performance Summary}

Table~\ref{tab:f1_summary} presents F1-scores for malicious node detection across all experimental conditions. These scores represent final detection accuracy after training convergence, computed from confusion matrices using trust threshold $\theta = 0.45$.

\begin{table}[htbp]
\centering
\caption{F1-Scores for Malicious Node Detection Across All Conditions}
\label{tab:f1_summary}
\begin{tabular}{|l|c|c|c|l|}
\hline
\textbf{Attack} & \textbf{RL} & \textbf{DRL} & \textbf{MARL} & \textbf{Best Agent} \\
\hline
NMA & 0.92 & \textbf{1.00} & 0.73 & DRL \\
CRA & 0.50 & 0.68 & \textbf{0.85} & MARL \\
AAA & 0.50 & \textbf{1.00} & \textbf{1.00} & DRL/MARL \\
BFI & \textbf{1.00} & \textbf{1.00} & \textbf{1.00} & All \\
TDP & 0.16 & 0.16 & 0.11 & None effective \\
\hline
\textbf{Average} & 0.62 & \textbf{0.77} & 0.74 & DRL \\
\hline
\end{tabular}
\end{table}

Several patterns emerge from this summary. DRL achieves the highest average F1-score (0.77), demonstrating that deep function approximation provides consistent advantages over tabular methods. However, MARL outperforms both alternatives against the Collusive Rumor Attack (0.85 vs. 0.68 for DRL and 0.50 for RL), validating our hypothesis that coordinated learning provides advantages against coordinated attacks. Most critically, all agents fail catastrophically against Time-Delayed Poisoning after sleeper activation, with F1-scores dropping to 0.11--0.16—effectively rendering the defense ineffective.

\subsection{Confusion Matrix Visualization}

\begin{figure}[htbp]
\centering
\includegraphics[width=0.9\textwidth]{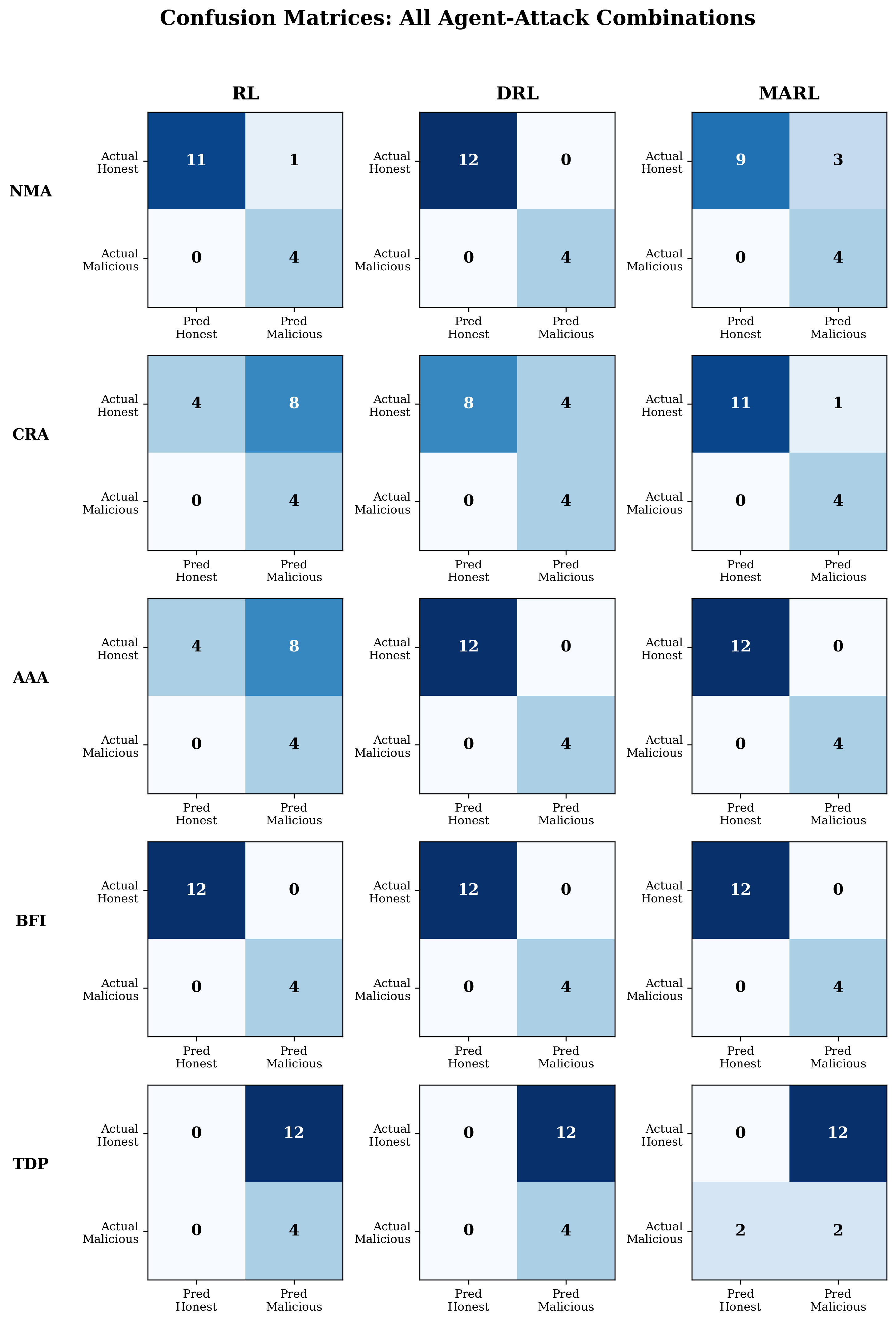}
\caption{Confusion matrices for all agent-attack combinations. Rows represent attacks (NMA, CRA, AAA, BFI, TDP); columns represent agents (RL, DRL, MARL). Color intensity indicates count magnitude. TDP row shows complete trust inversion with all honest nodes misclassified.}
\label{fig:confusion_grid}
\end{figure}

Figure~\ref{fig:confusion_grid} presents confusion matrices for all 15 experimental conditions in a consolidated grid format, enabling direct visual comparison across agents and attacks.

The visual pattern clearly distinguishes attack difficulty levels. The BFI row shows uniformly strong diagonal dominance across all agents, indicating effective detection. The CRA row reveals progressive improvement from RL to MARL, with decreasing off-diagonal (false positive) intensity. The TDP row presents a striking anomaly: the confusion matrices are effectively inverted, with high counts in the false positive cell and near-zero true negatives, indicating complete trust landscape inversion.

\subsection{Comparative Agent Performance}

\begin{figure}[htbp]
\centering
\includegraphics[width=0.9\textwidth]{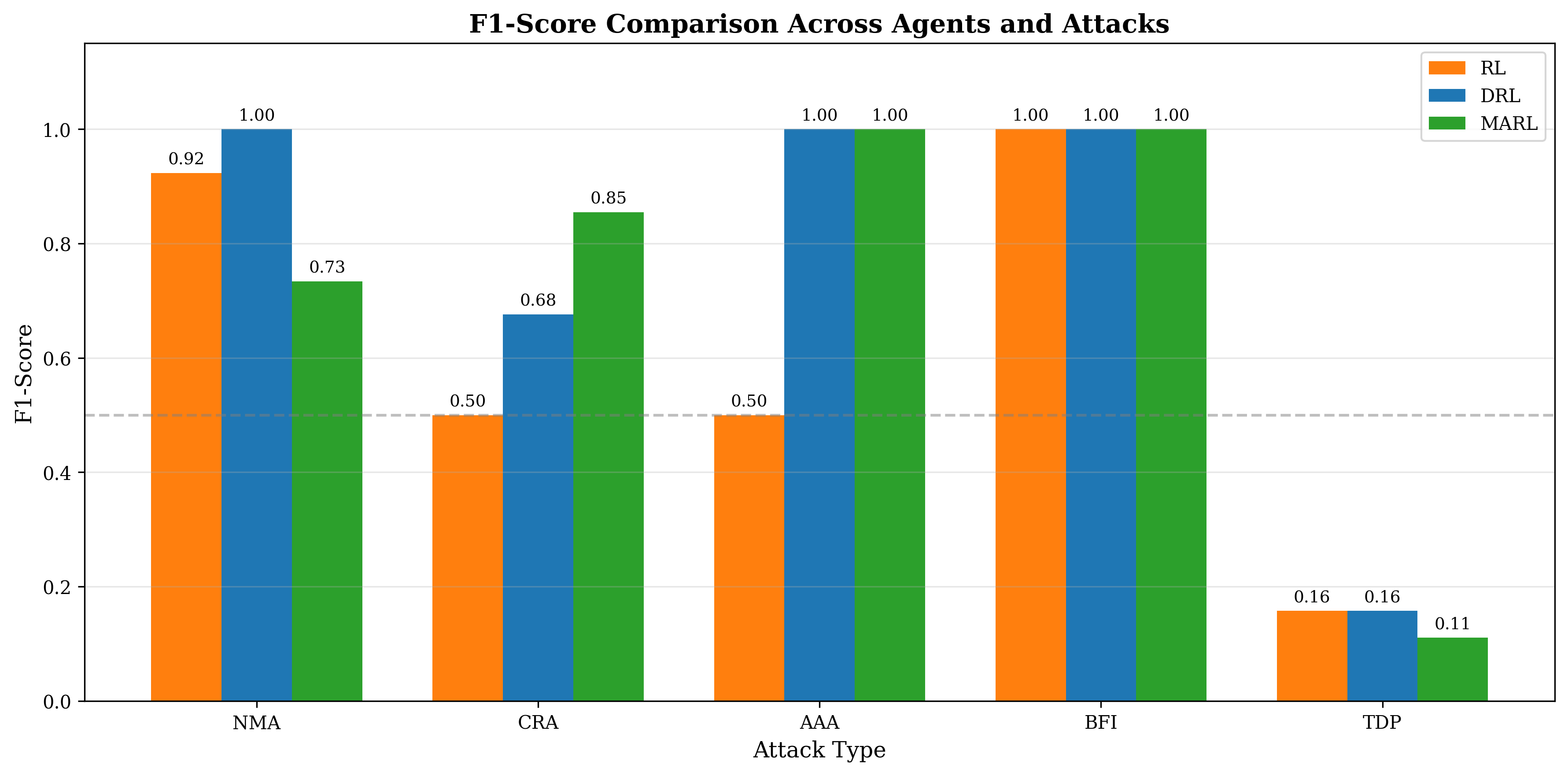}
\caption{F1-score comparison across agents for each attack type. MARL demonstrates clear advantage against CRA (coordinated attack), while DRL excels against NMA and AAA. All agents fail catastrophically against TDP after sleeper activation.}
\label{fig:f1_comparison}
\end{figure}

Figure~\ref{fig:f1_comparison} provides a bar chart comparison of F1-scores grouped by attack type, facilitating direct agent comparison within each adversarial scenario.

\subsubsection{RL Performance Analysis}

The tabular Q-learning agent exhibits inconsistent performance across attack types. It achieves perfect detection against BFI (F1=1.00) and strong performance against NMA (F1=0.92), but fails significantly against adaptive and coordinated threats. Against AAA, RL's F1-score drops to 0.50, indicating that the adaptive attack successfully exploits the limited representational capacity of tabular methods. The discretized state space loses critical information needed to distinguish attack-induced trust patterns from legitimate variations.

Against CRA, RL similarly achieves only F1=0.50. The coordinated trust manipulation creates state patterns that the discretized Q-table cannot adequately represent, leading to suboptimal policy decisions. These results confirm that simple RL approaches, while effective against unsophisticated threats, lack robustness against advanced adversaries.

\subsubsection{DRL Performance Analysis}

The Dueling Double DQN agent demonstrates strong overall performance, achieving perfect detection (F1=1.00) against NMA, AAA, and BFI. The continuous state representation and deep function approximation enable DRL to recognize subtle attack signatures that escape tabular methods. Against AAA specifically, DRL successfully counters the adaptive adversary despite its strategy rotation and defense evasion tactics, suggesting that the D3QN architecture learns robust features invariant to surface-level attack variations.

DRL shows improved but imperfect performance against CRA, achieving F1=0.68. While significantly better than RL (0.50), this falls short of MARL's 0.85. The coordinated trust inflation and suppression creates patterns that DRL's single-agent perspective cannot fully disentangle. When multiple malicious nodes simultaneously boost each other while penalizing an honest node, the resulting trust dynamics may appear similar to legitimate trust evolution from DRL's centralized viewpoint.

\subsubsection{MARL Performance Analysis}

The multi-agent framework exhibits distinctive strengths and weaknesses compared to single-agent approaches. MARL's standout result is against CRA, where it achieves F1=0.85—significantly outperforming DRL (0.68) and RL (0.50). The distributed learning architecture enables agents to recognize coordinated attack patterns from multiple perspectives, providing robustness against collusive behavior that single observers might miss.

Interestingly, MARL underperforms against NMA (F1=0.73 vs. DRL's 1.00). The uncoordinated nature of NMA may actually disadvantage MARL: with no attack coordination to detect, the multi-agent overhead introduces noise without corresponding benefit. This suggests that MARL's advantages are specific to coordinated threats rather than universal.

\subsection{Learning Dynamics}

Analysis of cumulative reward trajectories reveals distinct learning patterns across agent-attack combinations.

\subsubsection{Convergence Behavior}

\begin{figure}[htbp]
\centering
\includegraphics[width=\textwidth]{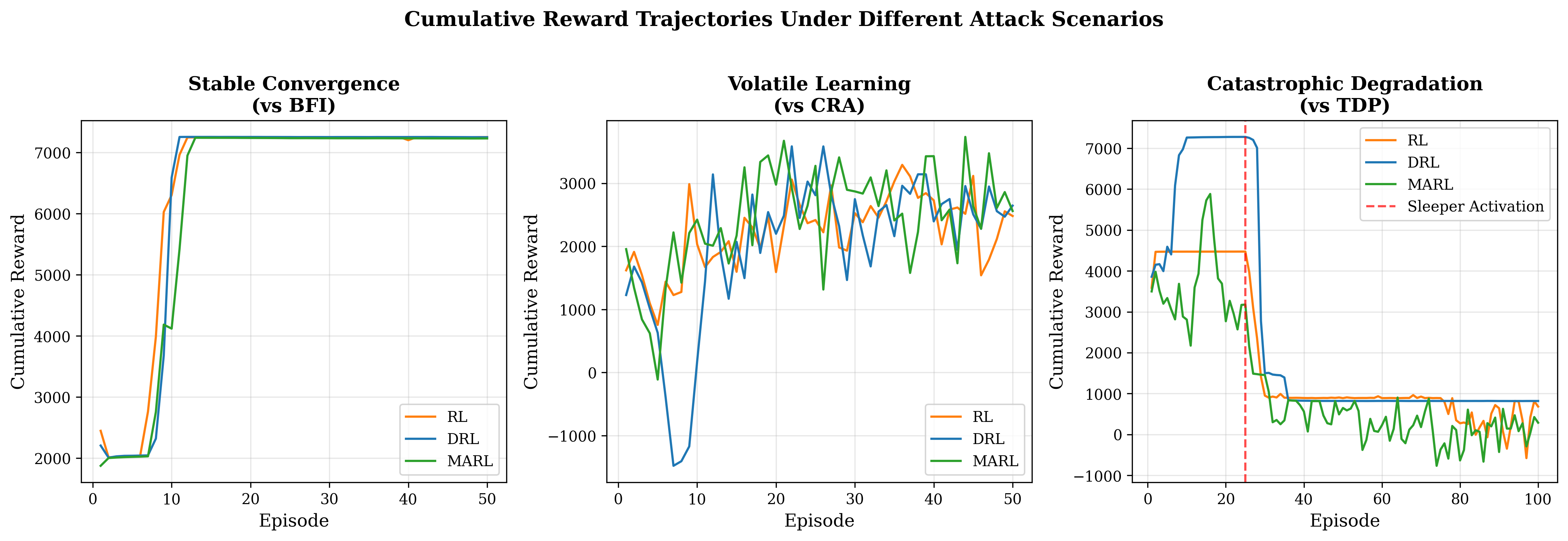}
\caption{Representative cumulative reward curves. Left: Stable convergence (vs BFI). Center: Volatile learning (vs CRA). Right: Catastrophic degradation (vs TDP, showing reward collapse at episode 25 when sleepers activate).}
\label{fig:reward_curves}
\end{figure}

Figure~\ref{fig:reward_curves} presents representative reward curves illustrating three qualitatively different learning outcomes: stable convergence, volatile non-convergence, and catastrophic degradation.

Against BFI and AAA, both DRL and MARL exhibit rapid convergence to stable high-reward policies, typically reaching asymptotic performance within 10--15 episodes. The reward curves show initial exploration phases with high variance, followed by exploitation phases with minimal fluctuation. This pattern indicates successful policy learning and robust defense establishment.

Against CRA, reward curves exhibit persistent volatility even after extended training. The coordinated attack continuously perturbs the trust landscape, preventing stable equilibrium. MARL shows somewhat reduced volatility compared to single-agent approaches, consistent with its superior F1 performance, but no agent achieves the stable convergence observed against other attacks.

\subsubsection{TDP Temporal Dynamics}

\begin{figure}[htbp]
\centering
\includegraphics[width=0.9\textwidth]{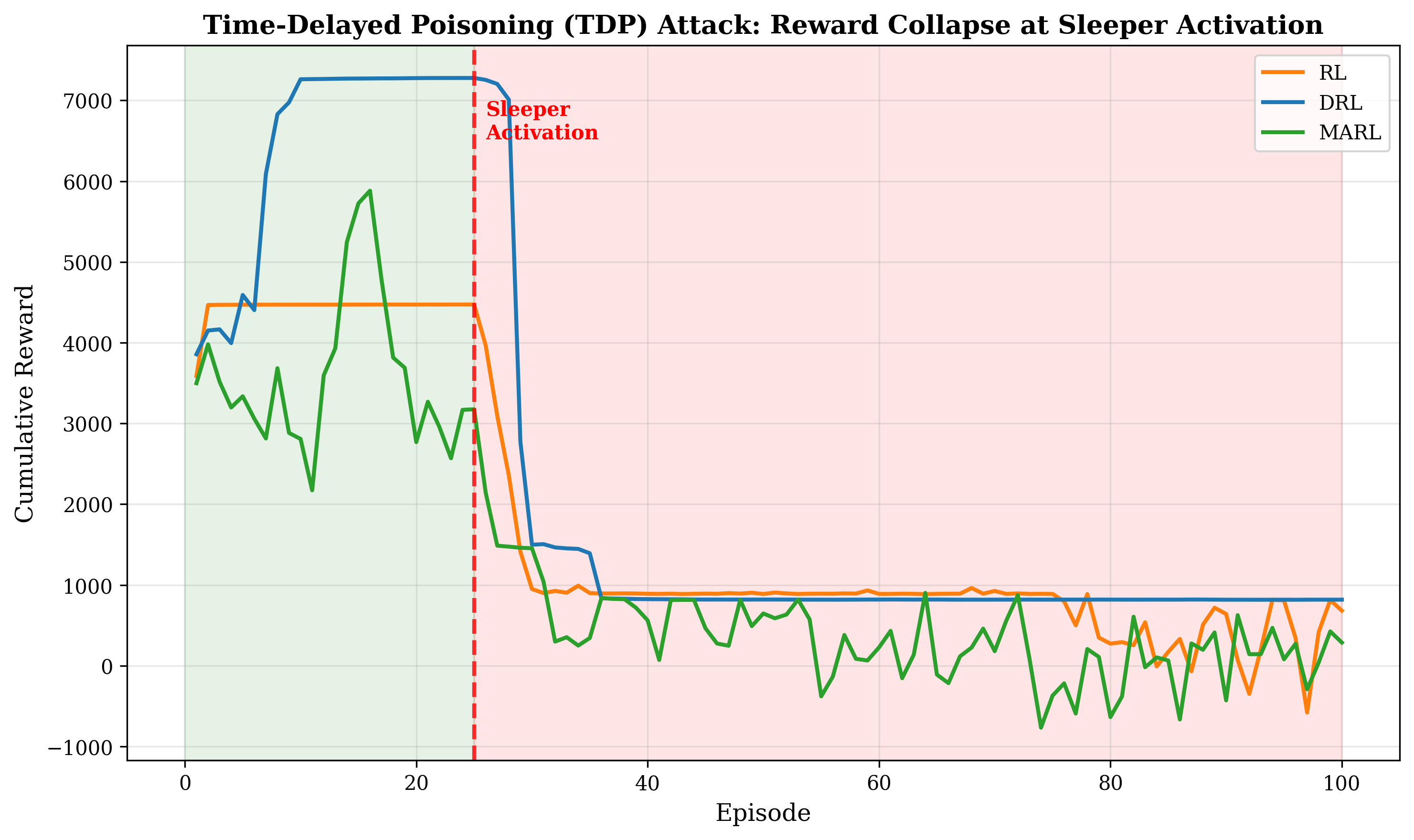}
\caption{Cumulative reward trajectories under TDP attack for all three agents. Vertical dashed line marks sleeper activation at episode 25. All agents experience catastrophic reward collapse, with no recovery observed through episode 100.}
\label{fig:tdp_dynamics}
\end{figure}

The TDP attack produces the most distinctive reward pattern, illustrated in Figure~\ref{fig:tdp_dynamics}. During the dormant phase (episodes 1--25), all agents achieve high rewards comparable to attack-free scenarios—the sleeper agents' honest behavior provides no signal for detection. At episode 25, reward curves show precipitous collapse from approximately 4000--6000 to near-zero or negative values, reflecting the sudden trust inversion caused by coordinated sleeper activation.

Critically, no agent demonstrates recovery capability post-activation. The reward curves remain suppressed through episode 100, indicating that the trust damage is effectively irreversible within our experimental timeframe. The resulting F1-scores of 0.11--0.16 represent near-complete detection failure. This result highlights a fundamental limitation of current trust mechanisms: systems optimized for responsiveness to recent behavior cannot defend against adversaries willing to invest in extended deception periods.

\subsection{F1-Score Evolution}

\begin{figure}[htbp]
\centering
\includegraphics[width=\textwidth]{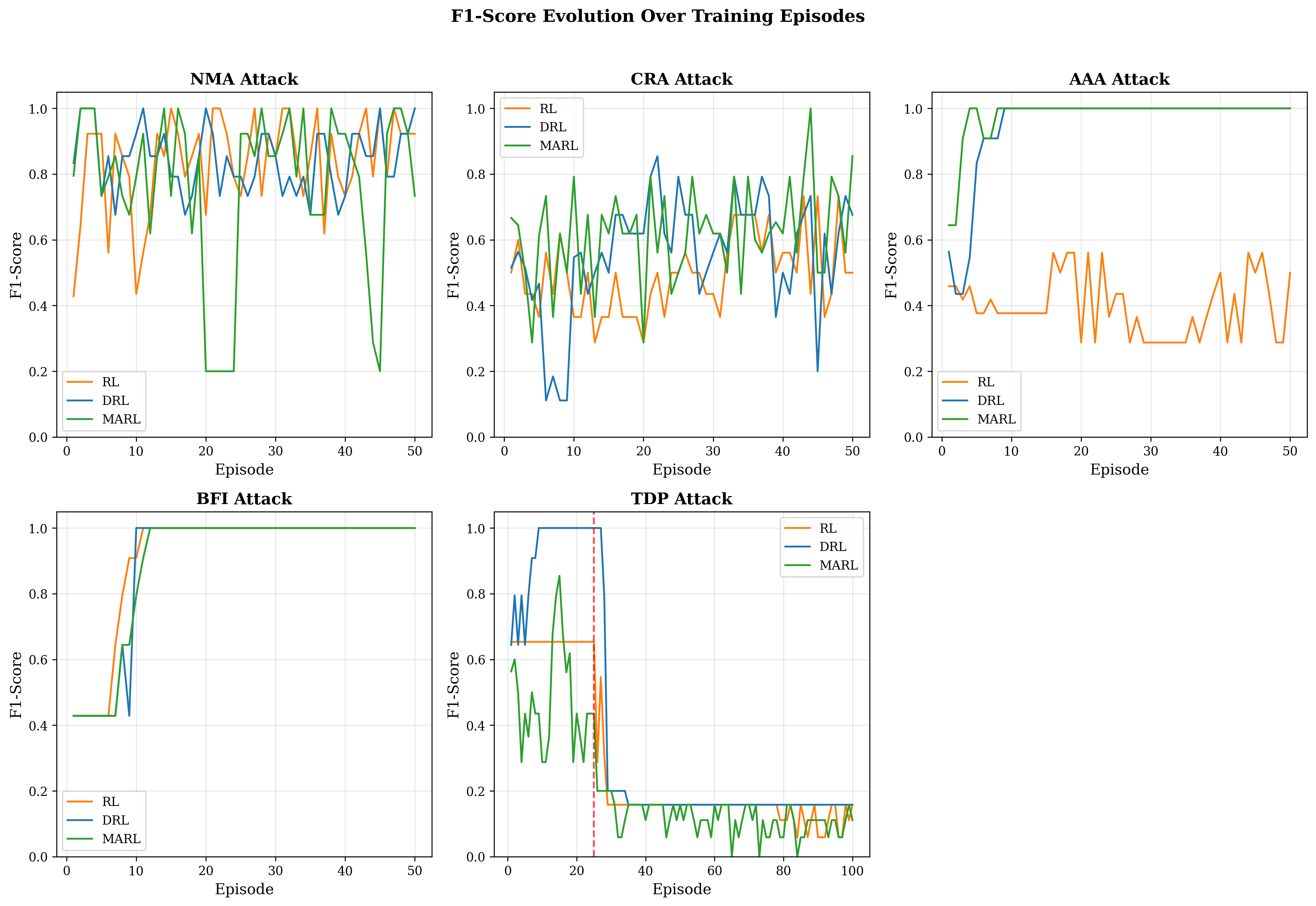}
\caption{F1-score evolution over training episodes for all five attack types. TDP shows high F1 during dormant phase followed by collapse at activation. CRA exhibits persistent instability. BFI and AAA converge rapidly for DRL/MARL.}
\label{fig:f1_evolution}
\end{figure}

Figure~\ref{fig:f1_evolution} shows how detection performance evolves during training across all attack scenarios.

The evolution patterns provide additional insight into agent learning dynamics. Against BFI, all agents achieve rapid F1 convergence to 1.0 within the first 10 episodes. Against NMA, DRL converges quickly while MARL shows slower, more variable improvement. The CRA curves reveal why MARL outperforms: while all agents show volatility, MARL maintains consistently higher F1 values throughout training.

The TDP evolution is particularly instructive. During episodes 1--25, all agents achieve F1 scores comparable to their performance against simpler attacks (0.8--1.0), providing false confidence in system security. The sudden collapse at episode 25 demonstrates the insidious nature of sleeper attacks—there is no gradual degradation to serve as warning.

\subsection{Key Findings Summary}

Our experimental evaluation yields several significant findings:

\textbf{Finding 1: Deep learning provides consistent advantages.} DRL achieves the highest average F1-score (0.77) across all attacks, demonstrating that continuous state representation and deep function approximation improve robustness over tabular methods (RL average: 0.62).

\textbf{Finding 2: Multi-agent learning excels against coordinated attacks.} MARL outperforms single-agent approaches specifically against CRA (F1=0.85 vs. 0.68 for DRL and 0.50 for RL), validating that distributed learning provides advantages when facing coordinated adversaries. This advantage does not transfer to uncoordinated attacks where MARL underperforms (NMA: 0.73 vs. DRL's 1.00).

\textbf{Finding 3: Adaptive attacks challenge simple methods but not deep learning.} RL fails against AAA (F1=0.50) while DRL and MARL maintain perfect detection (F1=1.00), indicating that sophisticated state representation is essential for countering adaptive adversaries that learn and rotate attack strategies.

\textbf{Finding 4: Time-delayed attacks defeat all current approaches.} No agent effectively defends against TDP post-activation, with F1-scores collapsing to 0.11--0.16. All honest nodes become misclassified while sleeper agents maintain high trust from their dormant-phase reputation. This represents a fundamental limitation requiring architectural innovations beyond the evaluated paradigms.

\textbf{Finding 5: Attack difficulty varies significantly.} BFI proves surprisingly manageable (all agents achieve F1=1.00) despite its theoretical sophistication, suggesting that trust-based consensus naturally resists Byzantine behavior through accumulated negative evidence. Conversely, temporal attacks exploiting trust accumulation dynamics pose the greatest threat, followed by coordinated collusion attacks.

Table~\ref{tab:findings_summary} summarizes agent recommendations by attack type based on our findings.

\begin{table}[htbp]
\centering
\caption{Agent Recommendations by Attack Type}
\label{tab:findings_summary}
\begin{tabular}{|l|l|l|}
\hline
\textbf{Attack Type} & \textbf{Recommended Agent} & \textbf{Rationale} \\
\hline
NMA (Naive) & DRL & Perfect detection (F1=1.00) \\
CRA (Collusive) & MARL & Best F1 (0.85), coordination advantage \\
AAA (Adaptive) & DRL or MARL & Both achieve perfect detection \\
BFI (Byzantine) & Any & All agents achieve F1=1.00 \\
TDP (Temporal) & None effective & Fundamental limitation (F1$\leq$0.16) \\
\hline
\end{tabular}
\end{table}

\section{Novelty and Contributions}
\label{sec:novel}
This study presents a unified, trust-aware blockchain framework that integrates Adaptive Fully Homomorphic Encryption (FHE), Attribute-Based Access Control (ABAC), and learning-based consensus control (RL/DRL/MARL) for adversarial IoT environments. Our contributions are:

\begin{enumerate}
    \item \textbf{Comprehensive Attack Taxonomy:} We formalize and implement five distinct attack families targeting trust-based blockchain consensus—Naive Malicious Attack (NMA), Collusive Rumor Attack (CRA), Adaptive Adversarial Attack (AAA), Byzantine Fault Injection (BFI), and Time-Delayed Poisoning (TDP)—spanning naive, coordinated, adaptive, Byzantine, and temporal threat models.

    \item \textbf{Adaptive Trust-Based Delegation:} We design a dynamic delegation mechanism where node trust evolves from on-chain behavioral feedback using Bayesian inference. Trust directly influences committee selection via Thompson Sampling, reducing overhead while preserving Byzantine fault tolerance.

    \item \textbf{FHE-Backed ABAC:} We couple ABAC with FHE so that access policies and attributes are evaluated over encrypted data, maintaining confidentiality during policy checks without trusted third parties and preventing adversaries from exploiting policy knowledge.

    \item \textbf{Attack-Aware Reward Design:} We formalize all five attack types within an RL setting, introducing a composite reward function that balances detection quality (F1-score), operational metrics (throughput), and explicit attack-specific penalties including collusion detection scores.

    \item \textbf{Systematic Comparison of RL Paradigms:} We provide the first systematic evaluation of tabular RL, Deep RL (Dueling Double DQN), and parameter-sharing MARL against five attack families under identical simulation conditions, isolating learning dynamics from experimental confounders.

    \item \textbf{Identification of Critical Vulnerability:} We demonstrate that Time-Delayed Poisoning attacks defeat all evaluated defense paradigms with F1-scores of 0.11--0.16, identifying a fundamental limitation of trust mechanisms optimized for behavioral responsiveness and motivating future research into temporal attack countermeasures.
\end{enumerate}

\section{Discussion}
\label{sec:discussion}

This section interprets our experimental findings in the broader context of blockchain IoT security, examines limitations of the current study, and discusses implications for real-world deployment.

\subsection{Interpretation of Results}

\subsubsection{The Coordination Hypothesis}

Our results provide strong evidence for what we term the \textit{coordination hypothesis}: defense mechanisms benefit from matching the coordination level of the attacks they face. MARL's superior performance against CRA (F1=0.85 vs. DRL's 0.68) demonstrates that distributed learning agents can recognize coordinated attack patterns that escape centralized observation. Conversely, MARL's underperformance against uncoordinated NMA (F1=0.73 vs. DRL's 1.00) suggests that multi-agent overhead becomes counterproductive when no coordination exists to detect.

This finding has practical implications for system design. Rather than deploying a single defense mechanism, adaptive systems might benefit from attack detection modules that identify coordination signatures and dynamically switch between single-agent and multi-agent defense modes. When coordination is detected, MARL provides measurable advantages; otherwise, DRL's lower overhead and higher precision make it preferable.

\subsubsection{The Representation Hypothesis}

The stark performance gap between tabular RL and deep learning approaches (average F1: 0.62 vs. 0.77) validates the \textit{representation hypothesis}: continuous state representations capture attack-relevant features that discretization destroys. This is particularly evident against AAA, where RL achieves only F1=0.50 while both DRL and MARL achieve perfect detection.

The Adaptive Adversarial Attack specifically targets defense mechanisms by learning and exploiting their patterns. Tabular RL's discretized state space creates predictable decision boundaries that AAA can probe and circumvent. Deep networks, by contrast, learn distributed representations where decision boundaries are implicit in high-dimensional weight spaces, making them inherently more difficult to reverse-engineer through behavioral observation.

\subsubsection{The Temporal Vulnerability}

The catastrophic failure against TDP (F1: 0.11--0.16 across all agents) reveals a fundamental tension in trust system design. Effective trust mechanisms must be responsive to behavioral changes—rewarding improvement and penalizing degradation. However, this responsiveness creates vulnerability to adversaries who invest in extended deception periods.

The TDP attack exploits this tension perfectly. During dormant phases, sleeper agents accumulate genuine positive trust evidence indistinguishable from honest behavior. Upon activation, they leverage accumulated reputation to inflict maximum damage before trust penalties accumulate. The trust mechanism's own reward structure becomes the attack vector.

This vulnerability is not specific to our implementation but inherent to any trust system that weights recent behavior. Potential countermeasures include behavioral consistency analysis (detecting sudden behavioral shifts), long-term trust memory (maintaining historical behavior profiles), and anomaly detection for coordinated activation patterns. However, these introduce their own trade-offs: reduced responsiveness to legitimate behavioral changes, increased storage requirements, and potential for honest nodes to be penalized for irregular but legitimate behavior patterns.

\subsection{Comparison with Prior Work}

Our results extend and partially contradict findings from prior studies. Muniswamy and Rathi \cite{muniswamy2024trust} reported high detection rates against CRA and NMA using D3QN, consistent with our DRL results. However, their evaluation did not include adaptive or temporal attacks, which we show pose significantly greater challenges. Our work demonstrates that detection success against basic attacks does not guarantee robustness against sophisticated adversaries.

Goh et al. \cite{goh_secure_2022} proposed trust-based delegation with DRL but evaluated only generic malicious behavior without distinguishing attack coordination levels. Our systematic comparison reveals that attack characteristics—particularly coordination and temporal dynamics—significantly influence defense effectiveness, suggesting that attack-agnostic evaluations may overestimate real-world security.

The MARL advantage we observe against coordinated attacks aligns with theoretical predictions from Standen et al. \cite{standen2025adversarial}, who argued that distributed defense mechanisms should provide advantages against distributed attacks. Our empirical confirmation of this hypothesis contributes concrete evidence to what was previously a theoretical argument.

\subsection{Limitations}

Several limitations constrain the generalizability of our findings.

\subsubsection{Simulation Environment}

Our experiments use a simulated blockchain environment with 16 nodes and 30\% malicious ratio. Real-world blockchain IoT deployments may involve thousands of nodes with varying malicious ratios over time. While our simulation captures essential trust dynamics, scale effects—including communication latency, partial observability, and emergent behaviors in large networks—remain unexplored. The computational cost of MARL scales with node count, potentially limiting its applicability in large-scale deployments.

\subsubsection{Attack Model Assumptions}

Our attack implementations assume adversaries with specific capabilities and strategies. Real attackers may combine attack elements (e.g., adaptive sleeper agents that learn optimal activation timing) or employ strategies not captured by our five-family taxonomy. The TDP activation episode (25) was fixed; adaptive activation based on accumulated trust or detected defense states could prove even more effective.

\subsubsection{Known Malicious Ratio}

Training assumes known ground truth labels for malicious nodes, enabling supervised reward computation. In deployment, ground truth is unavailable—nodes can only be classified based on observed behavior. This gap between training and deployment conditions may affect real-world performance, particularly for attacks designed to mimic honest behavior.

\subsubsection{Single-Objective Optimization}

Our reward function balances detection accuracy and operational metrics but does not explicitly optimize for energy efficiency, latency, or other constraints critical in resource-constrained IoT environments. Multi-objective formulations may reveal trade-offs not apparent in our single-objective evaluation.

\subsection{Implications for Deployment}

Despite limitations, our findings offer actionable guidance for practitioners deploying trust-based blockchain IoT systems.

\textbf{Agent Selection:} For environments where coordinated attacks are the primary concern (e.g., systems vulnerable to Sybil-based collusion), MARL provides measurable advantages despite higher computational overhead. For general-purpose deployment where attack type is unknown, DRL offers the best average performance (F1=0.77) with lower complexity than MARL.

\textbf{Temporal Attack Awareness:} Systems must acknowledge vulnerability to time-delayed attacks. Countermeasures might include periodic trust resets, behavioral consistency monitoring, or requiring sustained good behavior before granting elevated privileges. Our results suggest that no purely reactive defense suffices; proactive mechanisms for detecting dormant-phase positioning are essential.

\textbf{Defense Layering:} The FHE-secured ABAC component provides defense-in-depth by hiding policy logic from adaptive adversaries. While our experiments focused on learning agent performance, the combination of encrypted policy evaluation with adaptive trust management creates multiple barriers that adversaries must simultaneously overcome.

\textbf{Monitoring and Alerting:} The distinctive reward curve patterns we observed—stable convergence versus persistent volatility versus sudden collapse—suggest that runtime monitoring of defense agent behavior can provide early warning of attack presence. Sudden reward degradation, as observed in TDP activation, should trigger alerts and potentially activate fallback security measures.

\subsection{Ethical Considerations}

Our detailed documentation of attack mechanisms, particularly the highly effective TDP attack, raises dual-use concerns. We have chosen to publish this research because defensive innovation requires understanding attack capabilities, and the attacks we describe exploit fundamental properties of trust systems rather than specific implementation vulnerabilities. We encourage responsible disclosure practices and note that real-world deployment of such attacks would likely violate computer fraud laws in most jurisdictions.

\section{Conclusion}
\label{sec:conclusion}

This paper presented a comprehensive evaluation of reinforcement learning defenses for trust-based blockchain IoT consensus mechanisms against five distinct attack families. Our framework integrates Fully Homomorphic Encryption with Attribute-Based Access Control for privacy-preserving policy evaluation, combined with adaptive trust management using Bayesian inference and Thompson Sampling for delegate selection.

\subsection{Summary of Findings}

Our systematic comparison of tabular Q-learning (RL), Dueling Double Deep Q-Networks (DRL), and Multi-Agent Reinforcement Learning (MARL) against Naive Malicious Attacks (NMA), Collusive Rumor Attacks (CRA), Adaptive Adversarial Attacks (AAA), Byzantine Fault Injection (BFI), and Time-Delayed Poisoning (TDP) revealed significant performance variations:

\begin{itemize}
    \item DRL achieved the highest average F1-score (0.77) across all attacks, demonstrating that deep function approximation provides consistent advantages for malicious node detection.
    
    \item MARL outperformed alternatives against coordinated attacks (CRA: F1=0.85 vs. 0.68 for DRL), validating that distributed learning provides measurable advantages against distributed adversaries.
    
    \item All agents achieved perfect detection (F1=1.00) against Byzantine Fault Injection, suggesting that trust-based consensus naturally resists classical Byzantine behavior through accumulated negative evidence.
    
    \item Time-Delayed Poisoning proved catastrophic for all approaches, with F1-scores collapsing to 0.11--0.16 after sleeper activation. This finding identifies a critical vulnerability in trust mechanisms optimized for behavioral responsiveness.
\end{itemize}

\subsection{Contributions}

This work makes the following contributions to blockchain IoT security:

\begin{enumerate}
    \item A comprehensive attack taxonomy spanning naive, coordinated, adaptive, Byzantine, and temporal threat models, providing a systematic framework for evaluating trust-based defenses.
    
    \item The first systematic comparison of RL, DRL, and MARL for blockchain trust management under identical experimental conditions, enabling fair assessment of learning paradigm trade-offs.
    
    \item Empirical validation of the coordination hypothesis—that multi-agent defenses provide advantages specifically against coordinated attacks—with concrete performance measurements.
    
    \item Identification of Time-Delayed Poisoning as a critical vulnerability that defeats all evaluated defense paradigms, highlighting the need for temporal attack countermeasures in trust system design.
    
    \item A privacy-preserving framework combining FHE-secured ABAC with adaptive trust consensus, demonstrating that confidential policy evaluation can be integrated with learning-based defense.
\end{enumerate}

\subsection{Future Work}

Our findings motivate several directions for future research:

\textbf{Temporal Attack Countermeasures:} The catastrophic failure against TDP demands dedicated countermeasure research. Promising approaches include behavioral consistency analysis using change-point detection, long-term trust memory with decay-resistant historical profiles, and coordinated activation detection through temporal correlation analysis. Evaluating these mechanisms against adaptive sleeper agents that optimize activation timing presents a challenging but critical research direction.

\textbf{Hybrid Defense Architectures:} Our results suggest that different agents excel against different attack types. Future work should explore hybrid architectures that dynamically select or combine defense mechanisms based on detected attack characteristics, potentially achieving robust performance across the full attack spectrum.

\textbf{Scalability Analysis:} Extending evaluation to larger networks (hundreds to thousands of nodes) would assess whether our findings generalize to production-scale deployments. Of particular interest is whether MARL's coordination advantages persist as agent count increases or whether communication overhead eventually dominates.

\textbf{Transfer Learning:} Training separate models for each attack type is impractical in deployment where attack type is unknown. Transfer learning approaches that enable models trained on known attacks to generalize to novel attack patterns could significantly improve real-world applicability.

\textbf{Adversarial Robustness:} Our AAA implementation represents a first step toward modeling intelligent adversaries. More sophisticated adversarial training, where defense and attack agents co-evolve, could yield more robust defense mechanisms and more realistic attack models.

\subsection{Concluding Remarks}

Trust-based consensus mechanisms offer promising solutions for scalable blockchain IoT security, but their effectiveness depends critically on the sophistication of both defense mechanisms and anticipated attacks. Our evaluation demonstrates that while deep reinforcement learning provides significant advantages over traditional approaches, fundamental vulnerabilities remain—particularly against patient adversaries willing to invest in long-term deception.

The identification of Time-Delayed Poisoning as a critical threat highlights an important lesson: trust systems optimized for behavioral responsiveness inadvertently create attack surfaces for temporal exploitation. Addressing this vulnerability requires rethinking trust update dynamics, potentially sacrificing some responsiveness for temporal robustness.

As blockchain IoT systems become increasingly prevalent in critical infrastructure, supply chains, and autonomous systems, understanding and mitigating these vulnerabilities becomes essential. We hope this work contributes to that understanding and motivates continued research into robust, adaptive security mechanisms for decentralized systems.

\bibliographystyle{unsrt}  
\bibliography{references}  %%% Remove comment to use the external .bib file (using bibtex).
\end{document}